\documentclass[sigplan,10pt]{acmart}
\usepackage{hyperref}
\usepackage{epic,eepic}
\usepackage{graphics}
\usepackage{graphicx}
\usepackage{multicol}
\usepackage{xcolor}
\usepackage[vlined,ruled,linesnumbered,noresetcount]{algorithm2e}
\newcommand{\remove}[1]{}

\newcommand{\y}[1]{{\color{blue} #1}\normalcolor}

\newcommand{\yyy}[1]{{\color{green} #1}}

\newcommand{\danny}[1]{{\color{orange} #1}\normalcolor}
\newcommand{\ohad}[1]{{\textcolor{purple} {#1}}\normalcolor}
\newcommand{\lefteris}[1]{{\color{red} #1}\normalcolor}

\newcommand{\recovercode}[1]{{\textcolor{blue} {#1}}\normalcolor}
\newcommand{\flushcode}[1]{{\textcolor{red} {#1}}\normalcolor}

\newcommand{\hhere}[1]{{\bf [[#1]]}}
\newcommand{\here}[1]{}

\newcommand{\qedsymb}{\hfill{\rule{2mm}{2mm}}}

\SetKw{Continue}{continue}
\SetKw{OR}{or}
\SetKw{AND}{and}

\DontPrintSemicolon

\let\oldnl\nl
\newcommand{\nonl}{\renewcommand{\nl}{\let\nl\oldnl}}

\makeatletter
\newcommand{\removelatexerror}{\let\@latex@error\@gobble}
\makeatother

\newcommand{\ROptISB}{\mbox{\textsc{ROpt-ISB}}}

\newcommand{\search}{\mbox{\sc Search}}

\newcommand{\insertlst}{\mbox{\sc Insert}}
\newcommand{\delete}{\mbox{\sc Delete}}
\newcommand{\find}{\mbox{\sc Find}}

\newcommand{\init}{\mbox{$\bot$}}
\newcommand{\NULL}{\mbox{\sc Null}}

\newcommand{\CAS}{\mbox{\textit{CAS}}}
\newcommand{\True}{\mbox{\texttt{true}}}
\newcommand{\False}{\mbox{\texttt{false}}}

\newcommand{\AffectSet}{\mbox{\it AffectSet}}
\newcommand{\WriteSet}{\mbox{\it WriteSet}}
\newcommand{\NewSet}{\mbox{\it NewSet}}

\newcommand{\Info}{\mbox{Info}}
\newcommand{\pwb}{\mbox{\texttt{pwb}}}
\newcommand{\pfence}{\mbox{\texttt{pfence}}}
\newcommand{\psync}{\mbox{\texttt{psync}}}
\newcommand{\pbarrier}{\mbox{\texttt{pbarrier}}}
\newcommand{\barrier}{\mbox{\texttt{barrier}}}

\newcommand{\exInfo}{\mbox{ExInfo}}

\newcommand{\Key}{Key}

\newcommand{\NBBST}{LF-BST}

\newcommand{\func}[1]{\mbox{\sc #1}}

\newcommand{\comnospace}{\mbox{$\triangleright$}}
\newcommand{\com}{\mbox{\comnospace\ }}

\renewcommand{\paragraph}[1]{\smallskip\noindent{\bf #1}}

\startPage{1}

\setcopyright{none}

\usepackage{booktabs}   
\usepackage{subcaption} 

\begin{document}

\title[Tracking in Order to Recover]{Tracking in Order to Recover: Detectable Recovery of Lock-Free Data Structures}         

\author{Hagit Attiya}
\authornote{Supported in part by ISF grant 380/18.}
\orcid{1234-5678-9012}
\affiliation{%
  \institution{Technion}
	\country{Israel}
}
\email{hagit@cs.technion.ac.il}

\author{Ohad Ben-Baruch}
\authornotemark[2]
\affiliation{%
  \institution{Ben-Gurion University}
  \country{Israel}}
\email{ohadben@post.bgu.ac.il}

\author{Panagiota Fatourou}
\affiliation{%
	\institution{FORTH ICS \& \\ University of Crete, CSD}
	\country{Greece}}
\email{faturu@csd.uoc.gr}

\author{Danny Hendler}
\authornotemark[2]
\affiliation{%
	\institution{Ben-Gurion University}
	\country{Israel}}
\email{hendlerd@cs.bgu.ac.il}

\author{Eleftherios Kosmas}
\authornote{This research is co-financed by Greece and the European Union
(European Social Fund- ESF) through the Operational Programme «Human Resources Development,
Education and Lifelong Learning» in the context of the project “Reinforcement of Postdoctoral
Researchers - 2nd Cycle” (MIS-5033021), implemented by the State Scholarships Foundation (IKY).}
\affiliation{%
	\institution{University of Crete, CSD}
	\country{Greece}}
\email{ekosmas@csd.uoc.gr}

\begin{abstract}
This paper presents the \emph{tracking approach} for deriving
	\emph{detectably recoverable} (and thus also \emph{durable}) implementations
of many widely-used concurrent data structures.
Such data structures, satisfying \emph{detectable recovery},
are appealing for emerging systems featuring
byte-addressable \emph{non-volatile main memory} (\emph{NVRAM}),
whose persistence allows to efficiently resurrect
failed processes after crashes.
Detectable recovery ensures that after a crash,
every executed operation is able to recover and return a correct response,
and that the state of the data structure is not corrupted.

\emph{Info-Structure Based} (\emph{ISB})-tracking amends descriptor
objects used in existing lock-free helping schemes
with additional fields that track an operation's progress towards
completion and persists these fields to memory in order
to ensure detectable recovery.
ISB-tracking avoids full-fledged logging and tracks the progress
of concurrent operations in a \emph{per-process} manner,
thus reducing the cost of ensuring detectable recovery.

We have applied ISB-tracking to derive detectably recoverable 
implementations of a queue, a linked list, a binary search tree,  and
an exchanger.  Experimental results show the feasibility of the technique. 
\end{abstract}



\maketitle

\section{Introduction}

Byte-addressable \emph{non-volatile main memory} (\emph{NVRAM})
combines the performance benefits of conventional main memory
with the durability of secondary storage.
Systems with NVRAM 
are anticipated to be more prevalent in the near future.
The availability of durable main memory has increased the interest in
the \emph{crash-recovery} model, in which
failed processes may be resurrected after the system crashes.
Of particular interest is the design of \emph{recoverable concurrent
objects} (also called
\emph{persistent}~\cite{ChenQ-VLDB2015,CoburnCAGGJW-Asplos2011}
or \emph{durable}~\cite{VenkataramanTRC-FAST2011}),
whose operations can recover from crash-failures.
It is also important to be able to tell after recovery whether
an operation was executed to completion and if so,
what its response was, a property called \emph{detectable
recovery}~\cite{AttiyaBH-PODC2018,DBLP:conf/ppopp/FriedmanHMP18}.

In many computer systems (e.g., databases), detectable recovery is
supported by precisely \emph{logging} the progress of computations
to non-volatile storage, and replaying the log during recovery.
Logging imposes significant overheads in time and space.
This cost is even more pronounced for concurrent data structures,
where there is an extra cost of synchronizing accesses to the log.
Furthermore, replaying an operation in our setting, where some processes
may be concurrently recovering from crash-failures while others
have already completed recovery and proceed their normal execution,
requires to add new mechanisms to the original,
non-recoverable, implementation.

A key observation is that in the context of concurrent data structures,
full-fledged logging is not needed,
and the progress of an operation can be \emph{tracked} individually,
in a way supporting detectable recovery.
Moreover, \emph{many lock-free implementations
already encompass such tracking mechanisms},
which can be easily adapted to support detectable recovery.
This leads to the \emph{ISB-tracking approach} for designing recoverable
objects, based on explicitly maintaining an \emph{Info structure},
stored in non-volatile memory,
to track an operation's progress as it executes.
The Info structure allows a process to decide, upon recovery,
whether the operation's effect has already become visible
to other processes, in which case, the mechanism allows to
determine the response of the operation. 
(See Section~\ref{section:overview}.)

In many cases, ISB-tracking requires small changes to
the original code.
It significantly saves on the cost (in both time and memory)
incurred by tracking operations' progress,
by not having to track which instructions have been performed exactly,
but rather, specific stages of the operation.
Furthermore, even this can often be piggybacked on information
already tracked by lock-free concurrent data structures,
within \emph{operation descriptors}.
This means that operations efficiently maintain
and persist sufficient information for recovery,
and that the corresponding recovery code infers
whether the operation took effect before the failure,
in which case its response value is computed and returned.
These properties are what make our approach attractive.

ISB-tracking is widely applicable---it can be
used to derive recoverable versions of a large collection
of concurrent data structures.
We have applied it to derive a queue~\cite{MichaelS-PODC1996},
a linked list~\cite{DBLP:conf/wdag/Harris01} (Section~\ref{section:linked-list}),
a binary search tree~\cite{DBLP:conf/podc/EllenFRB10},
and an exchanger object~\cite{scherer2005scalable}.
The approach can be combined with a technique we call {\em direct-tracking}
which follows similar ideas as the log-queue~\cite{DBLP:conf/ppopp/FriedmanHMP18}
to get an elimination stack~\cite{DBLP:journals/jpdc/HendlerSY10}.

Detectability is a challenge even if caches
are non-volatile, i.e., writes are immediately persisted,
in program order.
However, ISB-tracking informs how {\em persistency instructions} (flushes and fences)
should be inserted for ensuring an implementation's correctness
in an efficient manner,
even when cache memories are volatile and their content is lost
upon a system-wide failure~\cite{DBLP:conf/wdag/IzraelevitzMS16}
(see Section~\ref{section:overview}).

We provide an experimental analysis (Section~\ref{section:evaluation})
to compare the performance of
ISB-tracking with all other generic schemes we are aware of
that can be used for deriving {\bf detectable} recoverable data structures.
The results show the feasibility of ISB-tracking. 
We have also implemented (in C++) hand-tuned, 
highly-optimized versions of detectably recoverable linked lists,
one using an existing general transformation~\cite{BBFW2019}, 
and another based on ideas from~\cite{DBLP:conf/ppopp/FriedmanHMP18}.

%
Experiments show that ISB-tracking performs better when contention is high.
For linked-lists, this is  because
ISB-tracking allows more precise insertion of persistency instructions,
based on a nuanced understanding of the consistency requirements
of the implementation.  

\remove{
Although there are many durable concurrent data structures and general schemes in the literature~\cite{+++},
the focus of our paper is on durable data structures that additionally satisfy {\bf detectability}.
Ensuring detectability comes with a cost and it would be unfair to compare the performance
of our detectable scheme with previous schemes that are durable but not detectable
(such as \cite{B-trees and other durable schemes pointed out by the reviewers+++}).
}

\remove{
Note that several other durable concurrent data structures and general schemes appear in the literature~\cite{Bteees, other citations mentioned by reviwewer}.
However, as ensuring detectability comes with a cost, we did not compare
the performance of our detectable scheme with previous schemes that are durable but not detectable~\cite{Bteees, other citations mentioned by reviwewer}.
}

Summarizing, the main contributions of this paper are:
\vspace*{-.12cm}
\begin{itemize}
	\item We propose ISB-tracking, a new mechanical transformation for deriving detectably recoverable implementations of concurrent data structures.
\item In a system with volatile caches, we present how persistency instructions can be added in ISB-tracking in a manner that
	enhances efficiency and scalability.
\item We apply ISB-tracking to get new detectably recoverable implementations of queues, linked lists, binary search trees, and exchangers.
\item We provide an experimental analysis to compare with {\em all} 
existing relevant transformations and detectably recoverable 
concurrent data structures we are aware of.
They show the feasibility of ISB-tracking and the 
good scalability it exhibits in many cases.
\end{itemize}




\section{System Model}
\label{section: Model}

We consider a system of asynchronous crash-prone \emph{processes} which
communicate through  \emph{base objects} supporting atomic read, write,
and Compare\&Swap (CAS) \emph{primitive} operations.

In the \emph{shared cache model},
the main memory is non-volatile, whereas
the data in the cache or registers are volatile.
Thus, primitive operations are applied to volatile memory,
and writes can be persisted to the non-volatile memory using explicit
flush instructions, or when a cache line is evicted.
Under \emph{explicit epoch persistency}~\cite{DBLP:conf/wdag/IzraelevitzMS16},
a write-back to persistent storage is triggered by a
\emph{persistent write-back (\pwb)} instruction.
The process may continue its execution after \pwb{}.
The order of \pwb{} instructions is not necessarily preserved.
When ordering is required,
a \pfence{} instruction orders preceding \pwb{} instructions
before all subsequent \pwb{} instructions.
A \psync\ instruction waits until all previous \pwb{} instructions
complete the write back.
For each location, persistent write-backs preserve program order.
We assume the \emph{Total Store Order} (\emph{TSO}) model,
supported by the x86 and SPARC architectures,
where writes become visible in program order.

The pseudocode for our generic technique
(Section~\ref{section:overview})
includes necessary persistency instructions,
showing how to efficiently persist information.
A simpler model is the
\emph{private cache model}~\cite{BBFW2019},
in which shared variables are always persistent.
In that 
model,
primitive operations are immediately applied
on the persistent memory,
and the state of each process is partitioned into
\emph{non-volatile private variables},
and \emph{local variables}
stored in \emph{volatile} processor registers.
In both models, private and local variables are accessed
only by the process they belong to.

At any point during the execution of an operation,
a system-wide \emph{crash-failure} (or simply a \emph{crash})
resets all volatile variables to their initial values,
while the values of (shared and local) non-volatile variables
are retained.\footnote{
In the private cache model, our approach and the derived algorithms work also in the more
general failure model where a process may fail individually.}
A process $q$ \emph{invokes} $Op$ to start its execution;
\emph{Op} \emph{completes} by returning a \emph{response value},
which is stored to a local variable of $q$.
The response value is always lost if a crash occurs before $q$ \emph{persists} it
(i.e., before it writes it to a non-volatile variable).

A \emph{recoverable} operation $Op$ has
an associated
\emph{recovery function}, denoted $Op.\texttt{Recover}$,
which the system calls when recovering $q$ after a system-failure
that occurred while it was executing $Op$.
Failed processes are recovered by the system asynchronously,
independently of each other;
the system may recover only a subset of these processes before another crash occurs.
The recovery code is responsible for finishing $Op$'s execution and returning its response.
An implementation is \emph{recoverable} if all its operations are recoverable.
Process $q$ may incur multiple crashes while executing $Op$
and $Op.\texttt{Recover}$,
so $Op.\texttt{Recover}$ may be invoked 
multiple times before $Op$ completes.
We assume that the system invokes $Op.\texttt{Recover}$ with the same
arguments as those with which $Op$ was invoked when the crash occurred.
For each process $q$,
we also use a non-volatile private variable $\text{CP}_q$,
that recoverable operations and recovery functions use for managing
check-points in their execution flow.\footnote{
    Some system support seems necessary for designing detectable
    data structures, and it has been assumed also in previous work~\cite{AttiyaBH-PODC2018,
    DBLP:conf/ppopp/FriedmanHMP18, DBLP:conf/podc/GolabH17}).}
When $q$ invokes a recoverable operation $Op$,
the system sets $\text{CP}_q$ to $0$ just before $Op$'s execution starts.
$\text{CP}_q$ can be read and written by recoverable operations
(and their recovery functions).
$\text{CP}_q$ is used by $q$ in order to persistently report
that the execution reached a certain point.
The recovery function can use this information in order
to correctly recover and to avoid re-execution of critical
instructions such as \CAS.

\remove{\y{We point out that our approach (as well as the recoverable
data structures we present) work even if we
assume that cache memories are volatile and their data are lost
every time a total-system failure occurs.}\danny{[[DH: I think the above text is problematic here. In general, our algorithms will work in such circumstances only if we apply flush operations in certain points, but our current model text abstracts away the issue of flushes. I think we would be better off without getting into this here.]]}} 

$Op$ is completed either directly or when,
after one or more crashes, the execution of the last instance of
$Op.\texttt{Recover}$ invoked for $q$ is complete.
In either case, $Op$'s response is written to a local variable of $q$.
We ensure
\emph{detectability}~\cite{DBLP:conf/ppopp/FriedmanHMP18}:
it is possible to determine, upon recovery,
whether the operation took effect, and its response value, if it did.

A recoverable implementation is \emph{wait-free} if,
when no crash occurs during the execution of a recoverable
operation (including its recovery function),
it completes in a finite number of steps.
It is \emph{lock-free} if, when no crash occurs,
\emph{some} recoverable operation completes in a finite number of steps.



\section{Info-Structure Based Tracking}
\label{section:overview}


Many lock-free implementations of data structures
(e.g.,~\cite{DBLP:conf/spaa/Barnes93,
DBLP:conf/podc/EllenFHR13,DBLP:conf/podc/EllenFRB10,
feldman2016efficient,DBLP:conf/ipps/WalulyaT17})
employ a \emph{helping mechanism} to ensure global progress,
even if processes crash.
They associate an information (\emph{Info}) structure
with each update, tracking the progress of the update
by storing sufficient information to allow its completion
by concurrent operations.

Our approach takes advantage of this \emph{Info-Structure-Based
(ISB)} helping to provide detectable recovery.
In brief, ISB helping works as follows:
An operation \textit{Op} by process $q$ initializes an Info structure
and then attempts to \emph{install} it in every node
that \textit{Op} is trying to change;
this is done by executing a \CAS\ to set a pointer in each of these nodes
to point to the Info structure initialized by \textit{Op}.
Once the Info structure is successfully installed,
$q$ continues the execution of \textit{Op} using the information stored in the Info structure.
Once the update completes, \textit{Op} \emph{invalidates}
the Info structure in all nodes pointed to it.
If it fails to install the Info structure on some node,
$q$ finds the Info structure of another operation installed at the node,
uses the information in it to complete the operation,
and then restarts \textit{Op}.

A key observation is that ISB helping goes a long way
towards making a data structure recoverable:
updates are idempotent and not susceptible to the \emph{ABA problem},
since they must ensure that an update is done exactly once,
even if several processes attempt to concurrently help it complete.
So, if the system crashes while $q$ is executing an operation \textit{Op},
upon recovery, $q$ can essentially re-execute $Op$
to completion by either using the information in the Info structure
for \textit{Op} (if it has already been installed)
or by starting from scratch.

To support detectability,
when $q$ recovers from a crash that occurred while executing
one of its operations \textit{Op}, its recovery code must be able to access the
Info structure $I$ of \textit{Op}.
This is achieved by allocating, for every process $q$,
a designated persistent \emph{recovery data} variable, $RD_q$,
that stores a reference to $I$.
Furthermore, a recovering process $q$ must be able to figure out
whether its failed operation took effect, and if it did,
what its response was.
To ensure this, a \textit{result} field is added to the
Info structure.
Process $q$, and every process helping \textit{Op},
sets the \textit{result} in $I$ before
invalidating $I$ from the relevant nodes.
Upon recovery, $q$ reads a reference to the last Info structure
from $RD_q$ and uses it to complete its last operation.
If the \textit{result} field of the Info structure is set,
the operation took effect, and $q$ returns its value.
Otherwise, \textit{Op} did not take effect and it can be restarted.
Even if \textit{Op} performed changes that have been later obliterated
by other operations,
the \textit{result} field of \textit{Op} would still be set.

\noindent
{\bf Detailed description.} 
Consider an implementation of a data structure that is represented
as a set of nodes, each with data fields and
pointers to other nodes in the data structure.
Each node $nd$ is augmented with an \textit{info} field containing a pointer
to an Info structure, which may be \emph{tagged}.
(We implement tagging by
setting the less significant bit of \textit{info}.)
When an Info structure
is first installed in $nd$ (i.e., $nd$'s \textit{info} field is set to point to this Info structure),
the node pointer to the Info structure is always tagged.
A node is {\em tagged} if its \textit{info} field is tagged.
Tagging a node acts like locking it. The node
may be later unlocked by untagging it.

High-level pseudocode appears in Algorithm~\ref{alg:ISB-generic}
(where the code in blue, dealing with recovery, and
the code in red, dealing with persistency, are explained below).
\emph{GetTagged} (\emph{getUntagged}) returns
a tagged (untagged) version of its argument
without changing its value.
We assume that \CAS{} returns the value it read from the variable.
An execution of an operation \textit{Op} by a process $q$ goes through one or more \emph{attempts},
each of which is an iteration of a while loop (Line~\ref{alg:ISB-while}),
until one of them is successful and \textit{Op} returns.
In each of its attempts, \textit{Op} first executes its {\em gather phase},
where it traverses the data structure
gathering those nodes that it will {\em affect},
i.e., those nodes that it will attempt to update or delete,
nodes that need to be locked for performing these updates and deletions,
as well as nodes that contain values the operation will return
or are needed to determine the operation's response.
As an example,
in a sorted linked list, a successful insert (or delete) affects the last two nodes it accesses
during its search.
On the other hand, a Find (or an unsuccessful update) affects only the last node it accesses.
This set of nodes is called the \emph{\AffectSet} of \textit{Op}.
Specifically, the \emph{\AffectSet}  is comprised of pairs, each containing a pointer to such a node
and the value of its \textit{info} field.

\begin{algorithm*}[tb]
	\nonl
	\begin{multicols*}{2}
		
	\removelatexerror
	\scriptsize
	
	\SetKwBlock{Begin}{}{}
	
	\begin{procedure}[H]
			\caption{() \small \func{Op} (args)}
			Info *\textit{opInfo} $:=$ \textbf{new} \Info\ () \label{alg1:opinfo1}\;
			\recovercode{$RD_q := \NULL$} \;
			\flushcode{\pbarrier\ ($RD_q$)} \;
			\recovercode{$CP_q := 1$} \;
			\flushcode{\pwb\ ($CP_q$)}; \flushcode{\psync} \label{alg:ISB-fence-init} \;
			
			\While {\True \label{alg:ISB-while}} {
				
				\Begin (\textbf{Gather Phase}) {
					Traverse the data structure gathering pairs $\langle \textit{nd}, \textit{ndinfo} \rangle$
					of pointers to those nodes (and to their \textit{info} fields)
					that \textit{Op} will affect, e.g. will try to update or delete, as well as nodes that contain the values the operation will return  (the \textit{info} field of each node
					is gathered on the first access to it)\;
				$\AffectSet{} :=$ list of all such pairs \;
				}
				\Begin (\textbf{Helping Phase}) {
					\uIf {there is a tagged \textit{ndinfo} field in a pair of \AffectSet} {
						\func{Help}(\textit{ndinfo}, \False) \label{alg:ISB-call-help} \;
						\Continue \;
					}
				}
				$\WriteSet :=$ list of structs of type Update containing those fields of nodes from \AffectSet{} that have to change, together with an old and a new value for each of them to perform the change using \CAS\;
				$\NewSet :=$ list of newly allocated nodes that are necessary to execute the operation, tagged with \textit{opInfo}\;
				$\textit{*opInfo} := \langle \func{Op}, \AffectSet, \WriteSet, \NewSet, \bot\rangle$ \tcp*{$\bot$ for result} \label{alg:ISB-new-opinfo}
                \flushcode{\pbarrier\ (\textit{*opInfo}, \NewSet)} \;
				\recovercode{$RD_q := \textit{opInfo}$} \;
				\flushcode{\pwb\ ($RD_q$)}; \flushcode{\psync} \label{alg:ISB-fence-after-install-info} \;
				\func{Help}$(\textit{opInfo}, \True)$ \label{alg:ISB-call-help-after-cas} \;
				\lIf {$\textit{opInfo}$$\rightarrow$$\textit{result} \neq \bot$} {
					\KwRet \textit{opInfo}$\rightarrow$\textit{result}
				}
			}
	\end{procedure}

	\begin{procedure}[H]
		\color{blue} {
			\caption{() \small \func{Op-Recover} (args)}
			
			Info *\textit{opInfo} := $\textit{RD}_q$ \;
			\lIf {$\textit{CP}_q = 0$ or $\textit{opInfo} = \NULL$} {
				Re-invoke \func{Op} (args)
			}
			\func{Help}(\textit{opInfo}, \True) \;
			\lIf {$\textit{opInfo}$$\rightarrow$$\textit{result} \neq \bot$} {
				\KwRet \textit{opInfo}$\rightarrow$\textit{result}
			}
			\lElse {
				re-invoke \func{Op}
			}
			
		}
	\end{procedure}

	\columnbreak

	\begin{procedure}[H]
		\caption{() \small \func{Help} (Info *\textit{opInfo}, boolean \textit{invoker})}
		
		\Begin (\textbf{Tagging Phase}) {
		\uIf {\textit{invoker} = \True} {
				$\langle$\textit{nd}, \textit{ndinfo}$\rangle :=$
        	    	    head of \textit{opInfo}$\rightarrow$\AffectSet
			}
			\lElse {
				$\langle$\textit{nd}, \textit{ndinfo}$\rangle :=$
           		     	second element of \textit{opInfo}$\rightarrow$\AffectSet
			}
%
			\While{\textit{nd} $\neq \NULL$} { \label{alg:ISB-fence-while}
				\textit{res} :=
                    \CAS(\textit{nd}$\rightarrow$\textit{info}, \textit{ndinfo}, \textit{getTagged}(\textit{opInfo})) \;
		\flushcode{\pwb\ (\textit{nd}$\rightarrow$\textit{info})} \;
				\uIf {\textit{res} $\neq$ \textit{ndinfo}
                    \AND \textit{res} $\neq$ \textit{getTagged}(\textit{opInfo})} {
					\lIf {\textit{invoker} = \True\ \AND $nd$ $\neq$ head element of \AffectSet} {
						\textit{opInfo} $:=$ \textbf{new} \Info\ ()  \label{alg1:opinfo2}	
					}
					\Begin (\textbf{Backtrack Phase}) {
							$\langle$\textit{nd}, \textit{ndinfo}$\rangle :=$
                                    previous element in \AffectSet \;
						\While {\textit{nd} $\neq \NULL$} {
							\CAS(\textit{nd}$\rightarrow$\textit{info}, \textit{getTagged}(\textit{opInfo}),
                                        \textit{getUntagged}(\textit{opInfo})) \;
                            \flushcode{\pwb\ (\textit{nd}$\rightarrow$\textit{info})} \;
							$\langle$\textit{nd}, \textit{ndinfo}$\rangle :=$
                                    previous element in \AffectSet \;
						}
						\flushcode{\psync} \;
						\KwRet \;
					}
				}
				$\langle$\textit{nd}, \textit{ndinfo}$\rangle :=$ next element in \AffectSet \;
			}
			\flushcode{\psync} \label{alg:ISB-fence-after-tagging} \;
		}
		\Begin (\textbf{Update Phase}) { \label{alg:ISB-fence-update-phase}
			\ForEach {Update structure \textit{st} in \WriteSet} {	
				perform the update by executing \CAS\ based on information contained in \textit{st} \;
				\flushcode{\pwb\ (updated field)} \;
			}
		}
		\recovercode{\textit{opInfo}$\rightarrow$\textit{result} $:=$ response
                of the operation described by \textit{opInfo}} \;
		\flushcode{\pwb\ (\textit{opInfo}$\rightarrow$\textit{result})}; \flushcode{\psync} \label{alg:ISB-fence-after-update} \;
		\Begin (\textbf{Cleanup Phase}) {
			\ForEach{node \textit{nd} in (\AffectSet{} $\cup$ \NewSet)
                        which is still part of the data structure} {
				\CAS(\textit{nd}$\rightarrow$\textit{info}, \textit{getTagged}(\textit{opInfo}),
					\textit{getUntagged}(\textit{opInfo})) \;
				\flushcode{\pwb\ (\textit{nd}$\rightarrow$\textit{info})}
			}
		\flushcode{\psync} \label{alg:ISB-fence-after-cleanup} \;
		}
	\end{procedure}

	\end{multicols*}

	\caption{ISB-Tracking (code for process $q$)}
	\label{alg:ISB-generic}
\end{algorithm*}

\textit{Op} then proceeds to its {\em helping phase}.
If an \textit{info} field \textit{ndInfo} (of a node \textit{nd}) in \AffectSet{}
is tagged, then \func{Help} is used to complete the operation that tagged the node
(i.e. the operation for which information is stored in \textit{ndInfo}),
before starting a new attempt.
After the helping phase,
the \WriteSet\ and the \NewSet---needed to complete \textit{Op}---are created.
The \WriteSet\ contains those fields of nodes from \AffectSet\ that need to change,
together with an old and a new value for each of them (needed to perform the change
using \CAS{}).
The \NewSet\ contains all newly allocated nodes by \textit{Op} that are
necessary for applying its updates (all these nodes are initially tagged
with a pointer to \textit{Op}'s \textit{opInfo}).
Then, the type of \text{Op}, its \AffectSet, its \WriteSet, its \NewSet, and the value $\bot$
(which is the initial value for the \textit{result} field)
are stored in the Info structure pointed to by \textit{opInfo}.
Next, \func{Help} is called with parameter \textit{opInfo}
to complete \textit{Op} itself.
If \func{Help} returns with $\textit{result} \neq \bot$,
then $Op$ has been performed and its result is returned;
otherwise, a new attempt is started.

\func{Help} tries to complete the operation described by \textit{opInfo}.
First, it applies CAS to try to install \textit{opInfo}
in every node of \AffectSet, in order ({\em tagging phase}).
If any of these CAS operations fails or
the associated \textit{Info} field is tagged by another operation,
then a \emph{backtrack phase} untags the nodes in \AffectSet{},
in reverse order.
After backtracking, \func{Help} returns.
If every node of \AffectSet{} is successfully tagged with \textit{opInfo},
then all changes to the \WriteSet\ are being performed
and the \textit{result} field is updated.
Finally, a \emph{cleanup phase} untags every node of \AffectSet{}
and \NewSet{}
 still in the data structure.

Note that a new Info structure is allocated only initially (Line~\ref{alg1:opinfo1}) and
after a non-empty backtrack phase (Line~\ref{alg1:opinfo2}). This avoids the
unnecessary overhead of allocating a new Info structure before calling \func{Help} in each attempt.
Note also that no process is aware of an operation $Op$ if its invoker does not tag at least one node.
Thus, given that tagging starts with the first node in the \AffectSet,
operations that help $Op$ do not need to try tagging that node. 
This is implemented with the second parameter of \func{Help} which differentiates the invoker of $Op$
by its helpers and starts tagging either from the first or the second element of $Op$'s \AffectSet,
respectively.

%


%

\emph{We make the following assumptions about the implementation:}
(a) It handles the \emph{ABA problem},
i.e., it does not store the same value into the same shared variable more
than once in any execution.
(b) The nodes in the \AffectSet{} or \WriteSet\ are always
accessed in the same order.\footnote{Many generic techniques assume a similar total order assumption
for ensuring lock-freedom and avoiding live-lock. The total order can be imposed in many ways
and is not necessarily fixed at the beginning of the execution.
For example, in a binary search tree, ordering can be determined by inorder or other
traversal orders.
Note also that it is only the concurrently active processes
that need to use the same total order, so different orderings
can be used during a single execution.}
(c) \func{Help} is \emph{idempotent} and can be executed
concurrently by several processes; \func{Help} is idempotent
if its changes are applied exactly once independently of how many
times they will be performed.


If \func{Help} completes the tagging phase for
the \textit{opInfo} of \textit{Op}, it returns only after
\textit{Op} takes effect:
all \CAS{} operations are applied to its \WriteSet,
its \textit{result} is updated and the cleanup phase is done.
If no process completes the tagging phase of \func{Help},
then \textit{Op} does not take effect.

In order to support detectable recovery,
a pointer to the Info structure being used in the last attempt of \textit{Op}
is persisted in $\textit{RD}_q$ (recall that $q$ is the process that invoked \textit{Op}).
Initially, $\textit{RD}_q$ is set to \NULL, and a check-point is set,
indicating that a new operation has started.
A pointer to the Info structure used in each attempt is stored in
$\textit{RD}_q$ before \func{Help} is called.
Upon recovery, \func{Op-Recover} is called with the same arguments as that of \textit{Op}
(see Section~\ref{section: Model}).
If the check-point is not set or $\textit{RD}_q$ is still \NULL,
\textit{Op} has made no changes and can simply be restarted.
Otherwise, the Info structure pointed to by $\textit{RD}_q$, \textit{opInfo},
indicates whether the last attempt of \textit{Op} was successful,
and if not, whether it crashed while making changes.
Since \func{Help} is idempotent,
recovery can call \func{Help}(\textit{opInfo}) to complete \textit{Op},
in case it is still in progress.
When \func{Help} returns, either \textit{Op} took effect
and \textit{result} stores its response,
or it did not and \textit{result} is $\bot$;
in the latter case, \textit{Op} can be re-invoked.
It is necessary to call \func{Help} first,
to deal with the case in which \textit{result} has been written but
the operation still needs to clean up,
in order to keep the data structure in a consistent state,
without nodes tagged by \textit{Op}.

\vspace{.1cm}
\noindent
{\bf Lock-freedom (proof outline).}
Since \func{Help} is wait-free, as it contains no unbounded loops, recovery is also wait-free.

If in an execution, all processes repeatedly perform failed
attempts, then the data structure remains static,
and only \textit{info} fields are changed.
Note that in this case, after some point, it must hold that
for every active operation \textit{Op}, \textit{Op}'s tagging phase does not complete.
By inspecting the pseudocode, we see that, otherwise,
\textit{Op}  would also complete successfully
(contradicting our hypothesis that all processes repeatedly perform failed attempts).

Consider a process $q$ making infinitely many failed attempts
while executing \textit{Op}.
Each attempt has a gather phase,
followed by a call to \func{Help} that fails  tagging
and performs a backtrack phase.
Since only \textit{info} fields are changed,
$q$ completes the gather phase with the same \AffectSet.
An attempt fails because some node in \AffectSet{} is tagged by
another operation $\textit{Op}'$.
If the process executing $\textit{Op}'$ does not take steps (i.e. it is {\em inactive}),
then the node is untagged, by $q$ or some other process,
during a backtrack phase in \func{Help}.
Thus, only a constant number of attempts fail because a node in
\AffectSet{} is tagged by an inactive operation.
Other attempts fail due to nodes in \AffectSet{} tagged by
operations that keep taking steps (i.e. they are {\em active}).

Eventually, we reach a scenario where all operations perform
failed attempts, each with the same \AffectSet,
in which one of the nodes is tagged by another active operation.
We now show that the total order assumption prevents such a scenario (of having a livelock).
Let $q'$ be the process that tagged the latest node in the
total order by which processes tag the nodes in their \AffectSet.
Clearly, no later node in \AffectSet{} of $q'$ is tagged by an active
operation, and thus $q'$ will complete its tagging phase,
and its attempt successfully.

\remove{
\ohad{We note that without the total store order a scenario of livelock is possible. Assume operation $\textit{Op}_1$ attempts to install info-structure in nodes $nd_1, nd_2$ according to this order, while operation $\textit{Op}_2$ attempts to install info-structure in nodes $nd_2, nd_1$ according to this order. Both operations may install info-structure on their first node in order, observe the second node to be marked by the other operation, and therefore backtrack by invalidating their info-structure. This scenario may repeat infinity many times, thus no operation makes progress.
We also note that the total store order can be determine during the execution, as long as all active processes agree on it. For example, by the structure of the data structure. This allows for a system with memory management to reuse nodes such that in different uses the node order is dynamically determined, and does not have to be set a priori.
}
}


\vspace{.1cm}
\noindent
{\bf Linearizability (proof outline).}
Intuitively, tagging puts \emph{``soft'' locks} on
the nodes in the \AffectSet{} of an operation \textit{Op} (initiated by a process $q$):
it ensures that these nodes 
are not updated (by operations not helping \textit{Op}) 
until \textit{Op}'s backtrack or cleanup phase completes.
The cleanup phase starts only after some process successfully updates
\textit{Op}'s \WriteSet\ and the response in \textit{result}.
Since \func{Help} is idempotent, a node is changed exactly once.
It can be shown that the \textit{info} field of a node does not hold
the same pointer twice whenever it is tagged.
Other fields are not prone to ABA, by assumption.
	
These features ensure that \func{Help} succeeds in updating the write
set, unless some other \func{Help} invocation for \textit{Op} has
started the cleanup phase, implying that the \WriteSet\ has already been updated.
Thus, \textit{Op} succeeds after its invocator $q$ collects a consistent
set of nodes, which do not change until some process completes the updates
on the \WriteSet.

We linearize \textit{Op} at the beginning of the update phase.
The above argument implies linearizability, as at this point the operation is guaranteed to complete, and other operations accessing any node in the \AffectSet{} must first help \textit{Op} to complete.
Moreover, detectability follows as well, since a cleanup phase is performed only after \textit{Op} completes all its updates, and \textit{result} is updated with \textit{Op}'s response.

\vspace*{.1cm}
\noindent
{\bf Persistency instructions for the shared cache model.}
There is a simple transformation~\cite{DBLP:conf/wdag/IzraelevitzMS16}
from the private cache to the shared cache model, which
puts a \pbarrier\ (a \pwb{} followed by a \pfence{})
after each access to a shared object,
and a \psync{} before returning from an operation.
However, the overhead of this transformation is very big
(see Section~\ref{section:evaluation}).

We now explain the customized persistence code, shown in red.
We note that an operation \textit{Op} tags news nodes it allocates with
a pointer to its Info structure,
before including them in its \NewSet{}.

After setting the check-point, allocating a new Info structure
and storing a reference to it in $RD_q$,
a \pwb{} followed by a \psync{} ensures that
the data is accessible upon recovery.
A \pbarrier{} after initializing $RD_q$ ensures that \pwb{}s
are executed in program order.
We also insert \pwb{} after every \CAS{} and write in \func{Help}.
A \psync{} at the end of every phase persists its changes
before the next phase.

An attempt to execute the changes of an operation \textit{Op},
with an Info structure \textit{opInfo}, happens after
tagging is complete and persisted in \func{Help}(\textit{opInfo}).
A crash before tagging ends may result in an old copy for the
\textit{Info} field of some nodes, but in this case,
no process started the update phase using the lost \textit{opInfo}.
Thus, upon recovery, the initiator of \textit{Op} will
call \func{Help}(\textit{opInfo}) to tag again.
Nodes are updated only after all nodes in \AffectSet,
as well as new nodes added to the data structure,
are tagged and persisted.
Every operation affected by these nodes,
must first complete \func{Help}(\textit{opInfo}).
A node is untagged only in the cleanup phase,
after all changes of the operation and its \textit{result} field
are persisted.

A crash during cleanup may cause an untagged node \textit{nd}
to be tagged, although another operation $\textit{Op}'$ might
have tagged \textit{nd} in the meantime.
Then the tagging phase by $\textit{Op}'$ is yet to be completed,
and both \textit{Op} and $\textit{Op}'$ invoke \func{Help}
on recovery.
Thus, \textit{Op} must first untag \textit{nd}
before $\textit{Op}'$ can tag it again.
%
To improve performance, all \pwb{} instructions can be issued
at the end of the phase, before the \psync{};
a single \pwb{} flushes all fields fitting in a cache line.

The correctness argument remains the same,
although the following scenario may occur:
a tagging phase (for \textit{opInfo}) may tag all the \AffectSet{},
but some of these tags (not necessarily in order)
were not persisted when a crash occurs.
At recovery, a different operation may already tagged some of
these untagged nodes, and re-tagging may fail.
Then, backtracking will only untag a prefix of \AffectSet{},
while other nodes are still tagged (for \textit{opInfo}).
A similar scenario occurs if a process fails during cleanup.
These scenarios do not violate durable linearizability
or lock-freedom:
Since \func{Help} is idempotent, any process that later
observes one of these tagged nodes, will fail during tagging
and untag the node.

\remove{
For the explicit epoch persistency model, the following changes are required.
All \pwb s in the body of \func{Op} must follow program order. For example, $RD_q$ must be persisted before $CP_q$ in Line~\ref{alg:ISB-fence-init}. A \pfence\ is added between the pairs of \pwb s to ensure that.
No change is required for the \func{Help} procedure.

\hhere{Reviewer1: The proof that the ISB technique provides durable linearizability and detectable execution is too informal.
The correctness section also needs improvements. The proof of linearizability should provide statements demonstrating that the proposed approach is durable linearizable. Additionally, the proof does not adequately explain how detectable execution is guaranteed. The proof statements should address how the recovery data (RD) and checkpoint (CP) represent the state of the operation's execution progress, and should reference specific line numbers when reasoning about correctness and progress.
(Ohad)}

}
\here{Y: Lefteris has some concerns about passing pointers as parameters in persistency instructions. Please discuss it and let me know the outcome.}

\noindent
{\bf Optimizing for read-only operations. }
%
%
We now discuss cases where ISB-tracking
can be optimized to achieve better performance.
Many concurrent data structures, 
including those implementing dictionaries,
support read-only operations (e.g., \func{Finds})
for which the \AffectSet{} contains just a single element;
moreover, they determine their response values based on
node fields that are immutable.
Under these conditions, 
ISB-tracking can be optimized so that a process $q$ executing such a read-only operation $Op$ is performed without
executing \func{Help} for $Op$, i.e. by skipping
the last three phases of Algorithm~\ref{alg:ISB-generic}.
The optimized pseudocode appears in Algorithm~\ref{alg:ROpt-ISB-generic}
(with changes appearing in green on Lines~\ref{alg2:opt1}-\ref{alg2:opt2} and~\ref{alg2:opt3}-\ref{alg:ROpt-ISB-ReadOnly-return-response}).
It checks whether the appropriate conditions hold for the operation
(i.e., if it is read-only and its \AffectSet{} contains a single element),
and if this is the case, it computes (Line~\ref{alg:ROpt-ISB-ReadOnly-compute-response})
and returns (Line~\ref{alg:ROpt-ISB-ReadOnly-return-response}) its response
(without calling \func{Help}). Note that the response, once computed, needs to be persisted.
To reduce the persistency cost, we compute the response earlier than the place we return it,
so it is persisted through the \pbarrier\ of Line~\ref{alg:ROpt-ISB-barrier-opinfo} and
the \psync{} of Line~\ref{alg:ROpt-ISB-fence-after-install-info}.

\begin{algorithm}[tb]
	\nonl
	\removelatexerror
	\scriptsize
	
	\SetKwBlock{Begin}{}{}
	
	\begin{procedure}[H]
			\caption{() \small \func{Op} (args)}
			Info *\textit{opInfo} $:=$ \textbf{new} \Info\ () \label{alg2:opinfo1}\;
			
			\recovercode{$RD_q := \NULL$} \;
			\flushcode{\pbarrier\ ($RD_q$)} \;
			\recovercode{$CP_q := 1$} \;
			\flushcode{\pwb\ ($CP_q$)}; \flushcode{\psync} \label{alg:ROpt-ISB-fence-init} \;
			
			\While {\True \label{alg:ROpt-ISB-while}} {
				
				\Begin (\textbf{Gather Phase}) {
					Traverse the data structure gathering pairs $\langle \textit{nd}, \textit{ndinfo} \rangle$ of
					pointers to those nodes (and to their \textit{info} fields)
					that \textit{Op} will affect, e.g. will try to update or delete, as well as nodes that contain the values the operation will return  (the \textit{info} field of each node
					is gathered on the first access to it)\;
					$\AffectSet{} :=$ list of all such pairs \;
				}
				\Begin (\textbf{Helping Phase}) {
					\uIf {there is a tagged \textit{ndinfo} field in a pair of \AffectSet} {
						\func{Help}(\textit{ndinfo},  \False) \label{alg:Opt-ISB-call-help} \;
						\Continue \;
					}
				}
				$\WriteSet :=$ list of Update structures containing those fields of nodes from \AffectSet{} that have to change, together with an old and a new value for each of them to perform the change using \CAS\;
				$\NewSet :=$ list of newly allocated nodes that are necessary to execute the operation, tagged with \textit{opInfo}\;
				$\textit{*opInfo} := \langle \func{Op}, \AffectSet, \WriteSet, \NewSet, \bot\rangle$ \tcp{$\bot$ for result} \label{alg:ROpt-ISB-new-opinfo}
				\yyy{\uIf {\textit{WriteSet} $= \emptyset$ \AND \textit{AffectSet} contains only one element} { \label{alg2:opt1}
					\recovercode{$\textit{opInfo}\rightarrow\textit{result} :=$ response \label{alg2:opt2}
					determined based on \textit{opInfo}} \label{alg:ROpt-ISB-ReadOnly-compute-response}
				} }
		                \flushcode{\pbarrier\ (\textit{*opInfo}, \NewSet)} \; \label{alg:ROpt-ISB-barrier-opinfo}
				\recovercode{$RD_q := \textit{opInfo}$} \;	
				\flushcode{\pwb\ ($RD_q$)}; \flushcode{\psync} \label{alg:ROpt-ISB-fence-after-install-info} \;
				\yyy{\uIf {\textit{WriteSet} $= \emptyset$ \AND \textit{AffectSet} contains only one element} { \label{alg2:opt3}
					\KwRet \textit{opInfo}$\rightarrow$\textit{result} \label{alg:ROpt-ISB-ReadOnly-return-response}
				} }
				\func{Help}(\textit{opInfo}, \True) \label{alg:ROpt-ISB-call-help-after-cas} \;
				\lIf {\textit{opInfo}$\rightarrow$\textit{result} $\neq \bot$} {
					\KwRet \textit{opInfo}$\rightarrow$\textit{result}
				}
			}
	\end{procedure}
	\remove{
	\begin{procedure}[H]
		\caption{() \small \func{Help} (Info *\textit{opInfo}, {\color{orange} boolean \textit{invoker}})}
		\Begin (\textbf{Tagging Phase}) {
			{\color{orange} \uIf {\textit{invoker} = \True} {
				$\langle$ \textit{nd}, \textit{ndinfo} $\rangle \leftarrow$
        	    	    head of \textit{opInfo}.\AffectSet
			}
			\lElse {
				$\langle$ \textit{nd}, \textit{ndinfo} $\rangle \leftarrow$
           		     	second element of \textit{opInfo}.\AffectSet
			}}
			{\bf lines~\ref{alg:ISB-fence-while}-\ref{alg:ISB-fence-after-tagging} of Algorithm~\ref{alg:ISB-generic} are repeated here}
		}
		{\bf lines~\ref{alg:ISB-fence-update-phase}-\ref{alg:ISB-fence-after-cleanup} of Algorithm~\ref{alg:ISB-generic} are repeated here}
	\end{procedure}  }
\caption{\ROptISB{}---ISB tracking optimized for read-only operations (code for process $q$)}
\label{alg:ROpt-ISB-generic}
\end{algorithm}

Section~\ref{section:linked-list} provides an example of a sorted linked list
which we obtain by applying Algorithm~\ref{alg:ROpt-ISB-generic}.
Note that optimizations aiming to improve the performance of
read-only operations may affect the way we assign linearization points.
For instance, in Algorithm~\ref{alg:ROpt-ISB-generic}, updates are still linearized at the beginning of their update phase.
However, a read-only operation $Op$ (that satisfies the optimization condition) is linearized
at the point that the Info field of the single
node added in the \AffectSet{} is read in $Op$'s last attempt.
To argue about correctness, we provide a simulation proof,
where for each execution $\alpha$ of Algorithm~\ref{alg:ROpt-ISB-generic},
we present a valid execution $\alpha'$ of Algorithm~\ref{alg:ISB-generic}
which contains the same operations as $\alpha$ and each operation
has the same response in both $\alpha$ and $\alpha'$.
Consider a read-only operation $Op$ (that satisfies the
optimization condition) executed by some process $q$ in $\alpha$.
We construct $\alpha'$ by letting $q$ execute {\em solo}  (i.e., without
any other process taking steps concurently with it), starting from the instruction
at which it is linearized, the part of
$Op$ (Algorithm~\ref{alg:ISB-generic})
that it is still to be executed in order to complete. Since
\textit{Op} does not change the data stucture,
the resulting execution (in which $Op$
tags and untags the single node in its \AffectSet{}
before any other process takes steps) is a valid execution
of Algorithm~\ref{alg:ISB-generic}. To get $\alpha'$, we apply this technique
to all read-only operations that can be optimized .

\section{Detectably Recoverable Linked List}
\label{section:linked-list}

\begin{algorithm}[tb]
	
	\nonl
	
	
	\SetKwBlock{Begin}{}{}	
		
	\removelatexerror
	\scriptsize

		

	\begin{procedure}[H]
		\caption{() \small boolean \insertlst\ (T $key$)}
		
		Node *\textit{newcurr} $:=$ \textbf{new} Node ($\init, \NULL, \NULL$) \; 
		Node *\textit{newnd} $:=$ \textbf{new} Node (\textit{key}, \textit{newcurr}, $\NULL$) \label{ROpt-ISB-ll-insert-create-newnd} \;
		Info *\textit{opInfo} $:=$ \textbf{new} \Info\ () \;
		
		\recovercode{$RD_q := \init$} \label{ROpt-ISB-ll-insert-set-RDp} \;
		\flushcode{\pbarrier\ ($RD_q$)} \;
		\recovercode{$\text{CP}_q := 1$} \tcp*{check-point; $RD_q$ is initialized} \label{ROpt-ISB-ll-insert-set-checkpoint}
		\flushcode{\pwb\ ($\text{CP}_q$)}; \flushcode{\psync} \;

		\While{\True}{
			\Begin (\textbf{Gather Phase} \tcp*[f]{search for right location to insert})  {	 \label{ROpt-ISB-ll-insert-gather-phase}
				$\langle$\textit{pred}, \textit{curr}, \textit{predInfo}, \textit{currInfo}$\rangle := \search(\textit{key})$  \label{ROpt-ISB-ll-insert-call-search} \;
				\lIf {\textit{curr}$\rightarrow$\textit{key} $=$ \textit{key}} {
					\AffectSet\ $:= \{\langle$\textit{curr},\textit{currInfo}$\rangle\}$
				}
				\lElse {
					\AffectSet\ $:= \{\langle$\textit{pred},\textit{predInfo}$\rangle,\langle$\textit{curr},\textit{currInfo}$\rangle\}$
				}
			}
			\Begin (\textbf{Helping Phase} \tcp*[f]{help other operations if necessary}) {	\label{ROpt-ISB-ll-insert-helping-phase}
				\uIf {\textit{isTagged}(\textit{predInfo})} {
					\func{Help} (\textit{predInfo}); \Continue \label{ROpt-ISB-ll-insert-help-pred}
				}
				\uElseIf {\textit{isTagged}(\textit{currInfo})} {
					\func{Help} ($currInfo$); \Continue \label{ROpt-ISB-ll-insert-help-curr}
				}
			}
			
			\textit{newcurr} $:=$ $\langle$\textit{curr} $\rightarrow$ \textit{key}, \textit{curr} $\rightarrow$ \textit{next}, \textit{getTagged}(\textit{opInfo})$\rangle$ \;
			\textit{newnd} $\rightarrow$ \textit{info} $:=$ \textit{getTagged}(\textit{opInfo}) \;
			\WriteSet $:= \{\langle$\textit{pred}$\rightarrow$\textit{next}, \textit{curr}, \textit{newnd}$\rangle\}$ \;
			\NewSet $:= \{$\textit{newnd}, \textit{newcurr}$\}$ \;
			*\textit{opInfo} $:= \langle$\insertlst, \AffectSet, \WriteSet, \NewSet $\init \rangle$ \label{ROpt-ISB-ll-delete-create-info} \;
			
			\yyy{\uIf (\tcp*[f]{\textit{key} in list}) {\textit{curr} $\rightarrow$ \textit{key} $=$ \textit{key}} { \label{ROpt-ISB-ll-insert-unsuccessful-if1}
				\recovercode{\textit{opInfo} $\rightarrow$ \textit{result} $:= \False$} \;	\label{ROpt-ISB-ll-insert-unsuccessful-write-response}
			}}
			\flushcode{\pbarrier\ (\textit{newcurr}, \textit{newnd}, \textit{*opInfo})} \;
			\recovercode{$RD_q$ $:=$ \textit{opInfo}} \tcp*{info for current attempt} \label{ROpt-ISB-ll-insert-update-RD}
			\flushcode{\pwb\ ($RD_q$)}; \flushcode{\psync} \;
			\yyy{\lIf (\tcp*[f]{\textit{key} in list}) {\textit{curr} $\rightarrow$ \textit{key} $=$ \textit{key}} { \label{ROpt-ISB-ll-insert-unsuccessful-if2}
				\KwRet \False \label{ROpt-ISB-ll-insert-return-false}
			}}
			\func{Help}(\textit{opInfo}, \True) \; \label{ROpt-ISB-ll-insert-help-opinfo}
			\lIf {\textit{opInfo} $\rightarrow$ \textit{result} $\neq \init$} {
				\KwRet \textit{opInfo} $\rightarrow$ \textit{result}
			}
		}
	\end{procedure}

			\begin{procedure}[H]
			\caption{() boolean \find\ (T $key$)}
			
			Info *\textit{opInfo} $:=$ \textbf{new} \Info\ () \;
			\While {\True} {
				\Begin (\textbf{Gather Phase}) {
					$\langle -, \textit{curr}, -, \textit{currInfo} \rangle := \search\ (\textit{key})$ \;
					\AffectSet $:= \{\langle$\textit{curr},\textit{currInfo}$\rangle\}$
				}
				\Begin (\textbf{Helping Phase}) {
					\uIf {IsTagged(\textit{currInfo})} {
						\func{Help}(\textit{currInfo}, \False) \;
						\Continue
					}
				}
				\yyy{\textit{result} $:=$ (\textit{curr} $\rightarrow$ \textit{key} $=$ \textit{key}) \;}
				\recovercode{\textit{opInfo} $\rightarrow$ \textit{result} $:=$ \textit{result}} \;
				\flushcode{\pbarrier\ (\textit{opInfo})} \;
				\recovercode{$RD_q$ $:=$ \textit{opInfo}} \;
				\flushcode{\pwb\ ($RD_q$)}; \flushcode{\psync} \;
				\yyy{\KwRet \textit{result}}
			}
		\end{procedure}
	


	\caption{Recoverable Linked List - \func{Find} and \insertlst{} (code for process $q$)}
	\label{alg:ROpt-ISB-ll-insert}
\end{algorithm}

\begin{algorithm}[tb]
	
	\nonl
	
	
	\SetKwBlock{Begin}{}{}	
		
	\removelatexerror
	\scriptsize


		\begin{procedure}[H]
			\caption{() \mbox{ \small \func{Help} (\Info\ *\textit{opInfo}, boolean \textit{invoker})}}
			
			\Begin (\textbf{Tagging Phase} ) {
				\uIf (\tcp*[f]{Try to tag \textit{pred}}) {\textit{invoker} $=$ \True } {
					{\color{black} \textit{result} $:=$ \CAS (\textit{opInfo} $\rightarrow$ \textit{pred} $\rightarrow$ \textit{info}, \textit{opInfo} $\rightarrow$ \textit{predInfo}, \textit{getTagged}(\textit{opInfo})) \tcp*{\textbf{tag} \CAS} \label{ROpt-ISB-ll-HelpOp-pred-tag-CAS}
					\flushcode{\pwb\ (\textit{opInfo} $\rightarrow$ \textit{pred} $\rightarrow$ \textit{info})} \;}
					{\color{black}\uIf 
									{\textit{result} $\neq$ \textit{opInfo} $\rightarrow$ \textit{predInfo} \AND \textit{result} $\neq$ \textit{getTagged}(\textit{opInfo})} {
						\KwRet \label{ROpt-ISB-ll-HelpOp-pred-tag-CAS-failed-return}
					}}			
				}
				
				\tcc{Try to tag \textit{curr} for removal}
				\textit{result} $:=$ \CAS (\textit{opInfo} $\rightarrow$ \textit{curr} $\rightarrow$ \textit{info}, \textit{opInfo} $\rightarrow$ \textit{currInfo}, \textit{getTagged}(\textit{opInfo})) \tcp*{\textbf{tag} \CAS} \label{ROpt-ISB-ll-HelpOp-curr-mark-CAS}
				\flushcode{\pwb\ (\textit{opInfo} $\rightarrow$ \textit{curr} $\rightarrow$ \textit{info})} \;
				\uIf (\tcp*[f]{\textit{opInfo} $\rightarrow$ \textit{curr} is not tagged}) {\textit{result} $\neq$ \textit{opInfo} $\rightarrow$ \textit{currInfo} \AND \textit{result} $\neq$ \textit{getTagged}(\textit{opInfo})} {
					\lIf {\textit{invoker = \True}} { \textit{opInfo} $:=$ \textbf{new} \Info\ ()  \label{ROpt-ISB-ll-HelpOp-new-opinfo} }
					\Begin (\textbf{Backtrack Phase}) {
						\CAS (\textit{opInfo} $\rightarrow$ \textit{pred} $\rightarrow$ \textit{info}, \textit{getTagged}(\textit{opInfo}), \textit{getUntagged}(\textit{opInfo}) \; \label{ROpt-ISB-ll-HelpOp-pred-untag-CAS}
						\flushcode{\pwb\ (\textit{opInfo} $\rightarrow$ \textit{pred} $\rightarrow$ \textit{info})}; \flushcode{\psync\ ()} \;
						\KwRet \tcp*{operation attempt failed}
					}
				}
				\flushcode{\psync\ ()} \;
			}
			\Begin (\textbf{Update Phase} \tcp*[f]{update \textit{pred} $\rightarrow$ \textit{next}}) {
				\ForEach {$\langle\textit{ptr},\textit{oldVal},\textit{nval}\rangle$ in \WriteSet} {	

					\CAS(\textit{ptr}, \textit{oldVal}, \textit{newVal}) \label{ROpt-ISB-ll-CompleteTagged-CAS} \;
					\flushcode{\pwb\ (ptr)} \;
				}
			}
			\tcc{complete operation}
			\recovercode{\textit{opInfo} $\rightarrow$ \textit{result} $:=$ \True} \tcp*{update response} \label{ROpt-ISB-ll-CompleteTagged-set-result}
			\flushcode{\pwb\ (\textit{opInfo} $\rightarrow$ \textit{result})}; \flushcode{\psync\ ()} \;
			
			\Begin (\textbf{Cleanup Phase}) {
				\ForEach{node \textit{nd} in (\AffectSet{} $\cup$ \NewSet)} {
				\CAS(\textit{nd}$\rightarrow$\textit{info}, \textit{getTagged}(\textit{opInfo}),
					\textit{getUntagged}(\textit{opInfo})) \;
				\flushcode{\pwb\ (\textit{nd}$\rightarrow$\textit{info})}
			}
		\flushcode{\psync} \label{alg:ROpt-ISB-fence-after-cleanup} \;

			}
			\KwRet \label{ROpt-ISB-ll-CompleteOp-return-true}
					
		\end{procedure}
	\caption{Recoverable Linked List - auxiliary functions (code for process $q$)}
	\label{alg:ROpt-ISB-ll-help}
\end{algorithm}

In this section, we illustrate how to apply Algorithm~\ref{alg:ROpt-ISB-generic} 
to get a detectable linked list. 
The list is sorted in increasing order of keys,
with two sentinel nodes, \emph{head} and \emph{tail},
holding keys $-\infty$ and $+\infty$. 
The \textit{next} field of a node points to the next node in the list.
A node $nd$ may be tagged either for update
(indicating its \textit{next} field is about to change),
in which case it is untagged after the update completes,
or for deletion (indicating it is to be deleted),
in which case it remains tagged forever.
When $nd$ is tagged, its Info structure contains information
necessary to complete the operation that tagged $nd$.
A field \textit{opType} in the Info structure indicates the operation
type (\insertlst\ or \delete).
(More details, including data types, shared variables, and initialization values of the algorithm 
are provided in the supplementary material.)

An instance $Op$ of \insertlst$(k)$ (Algorithm~\ref{alg:ROpt-ISB-ll-insert}),
executed by a process $q$, 
calls \search\  during its gather phase 
(Lines~\ref{ROpt-ISB-ll-insert-gather-phase}-\ref{ROpt-ISB-ll-insert-call-search}) 
to get pointers \textit{pred} and \textit{curr} to the nodes
between which $k$ should be added,
and their \emph{info} fields. If $Op$ is successful, these are the nodes 
contained in $Op$'s \AffectSet. Thus, the helping phase
(Lines~\ref{ROpt-ISB-ll-insert-helping-phase}-\ref{ROpt-ISB-ll-insert-help-curr}) 
simply checks whether these two nodes are tagged and calls \func{Help}
if needed. 

If the key $k$ to be inserted is already in the list,
$Op$ is read-only and behaves like a \func{Find}.
So, in this case, the \AffectSet{} includes
just the last node accessed by its search. We can then apply the
optimization for read-only operations,
as shown in Lines~\ref{ROpt-ISB-ll-insert-unsuccessful-if1}-\ref{ROpt-ISB-ll-insert-unsuccessful-write-response} and 
\ref{ROpt-ISB-ll-insert-unsuccessful-if2}-\ref{ROpt-ISB-ll-insert-return-false}.
Otherwise, $Op$ calls \func{Help} (Line~\ref{ROpt-ISB-ll-insert-help-opinfo}),
shown in Algorithm~\ref{alg:ROpt-ISB-ll-help}. 
\func{Help} starts by executing $Op$'s tagging phase: 
It first tries to tag \textit{pred} for update
(Line~\ref{ROpt-ISB-ll-HelpOp-pred-tag-CAS}),
so that it can later change its \textit{next} field
(Line~\ref{ROpt-ISB-ll-CompleteTagged-CAS}).
(\emph{IsTagged} checks whether the \textit{Info}
field of a node is tagged.)

Next, \func{Help} tries to tag \textit{curr} for deletion
by a {\em mark} \CAS{} (Line~\ref{ROpt-ISB-ll-HelpOp-curr-mark-CAS}).
If this \CAS{} fails, $q$ untags $pred$
(Line~\ref{ROpt-ISB-ll-HelpOp-pred-untag-CAS})
and restarts its own operation.
Once both nodes are tagged, \insertlst{} is guaranteed to succeed.
The new node containing $k$ is inserted between \textit{pred}
and the copy of \textit{curr}, by a \CAS{} that
points the \textit{next} field of \textit{pred} to the new node
(Line~\ref{ROpt-ISB-ll-CompleteTagged-CAS}).
A copy of \textit{curr} is used to avoid the ABA problem.
A successful \insertlst\ allocates two new nodes, as well as
a new Info object only each time it fails to tag $curr$
(Line~\ref{ROpt-ISB-ll-HelpOp-new-opinfo}).

\delete\ is simpler than \insertlst\,
since it does not allocate new nodes;
its code appears as Algorithm~\ref{alg:ROpt-ISB-ll-delete}
in the supplementary material.
$\delete(k)$ uses \search\ to get \textit{pred} and \textit{curr},
and then tries to tag \textit{pred} for update
(Line~\ref{ROpt-ISB-ll-HelpOp-pred-tag-CAS})
and \textit{curr} for deletion
(Line~\ref{ROpt-ISB-ll-HelpOp-curr-mark-CAS}).
If both nodes are tagged, \delete{} is guaranteed to succeed.
The node with key $k$ is physically deleted from the list
with a \CAS\ (Line~\ref{ROpt-ISB-ll-CompleteTagged-CAS}).

$\func{Find}(k)$ (Algorithm~\ref{alg:ROpt-ISB-ll-insert}) is read-only and computes its response based on immutable fields. 
Moreover, $Op$'s \AffectSet{} contains just the node pointed by $curr$. 
Therefore, \find{} is optimized to avoid installing
an Info structure.
Recovery is achieved in exactly the same way as in Algorithm~\ref{alg:ROpt-ISB-generic}.

\section{Evaluation}
\label{section:evaluation}

\newcommand{\isb}{\mbox{\textsc{Isb}}}
\newcommand{\ISB}{\mbox{\textsc{Isb}}}
\newcommand{\ISBOpt}{\mbox{\textsc{Isb-Opt}}}
\newcommand{\capsule}{\mbox{\textsc{Capsule}}}
\newcommand{\direct}{\mbox{\textsc{Direct-Tracking}}}
\newcommand{\clflush}{\mbox{\texttt{clflush}}}
\newcommand{\mfence}{\mbox{\texttt{mfence}}}
\newcommand{\DT}{\mbox{\textsc{DT}}}
\newcommand{\DTOpt}{\mbox{\textsc{DT-Opt}}}
\newcommand{\CP}{\mbox{\textsc{Capsules}}}
\newcommand{\Capsules}{\mbox{\textsc{Capsules}}}
\newcommand{\CapsulesOpt}{\mbox{\textsc{Capsules-Opt}}}
\newcommand{\CPIZ}{\mbox{\textsc{Capsules-IZ}}}
\newcommand{\HarrisLL}{\mbox{\textsc{Harris-LL}}}

\newcommand{\CPG}{\mbox{\textsc{Capsules-General}}}
\newcommand{\CPN}{\mbox{\textsc{Capsules-Normal}}}
\newcommand{\MSQUEUE}{\mbox{\textsc{MS-Queue}}}

\noindent
{\bf Evaluated Implementations.}
For our experiments, we use the Harris' linked list~\cite{DBLP:conf/wdag/Harris01,DBLP:books/daglib/0020056} as our example data structure. 
We compare our general approach (described in Algorithm~\ref{alg:ROpt-ISB-generic}) 
with capsules~\cite{BBFW2019}. 
Capsules is a syntactical transformation which can be applied
to concurrent algorithms that use only read and \CAS{} operations
to make them detectably recoverable. They achieve this 
by partitioning their code into \emph{capsules},
each containing a single \CAS{} operation,
and replacing each \CAS{} with its recoverable version~\cite{AttiyaBH-PODC2018}.
In general, a single operation may be partitioned to multiple capsules, but \emph{normalized} implementations~\cite{TimnatP-PPoPP2014} can be
optimized so that each operation is partitioned to only two capsules.
We experimented with both the general and the normalized variants. Since the normalized variant consistently outperformed the general variant, we only present the results of the former. 
To appropriately add persistency instructions to capsules without jeopardizing the generality of the approach,
it is proposed in~\cite{BBFW2019} to use a
general durability transformation~\cite{DBLP:conf/wdag/IzraelevitzMS16} (which adds \pwb\ and \pfence\  after each access to shared memory).

We compare the detectably recoverable linked list of Section~\ref{section:linked-list}, which we call \ISB,
with a linked list implementation, called \Capsules, obtained by applying the \emph{capsules} transformation
(plus the durability transformation of~\cite{DBLP:conf/wdag/IzraelevitzMS16}), to Harris' linked
list~\cite{DBLP:conf/wdag/Harris01,DBLP:books/daglib/0020056}.
We are not aware of other general schemes for deriving detectably recoverable data structures
in the literature. As ensuring detectability is costly, we did not 
compare the performance of \ISB\ with other schemes 
(e.g.~\cite{chatzistergiou2015rewind,DBLP:journals/pacmpl/CohenFL17,IKK2016,LIL2018,VTS2011,CorreiaFR20}) 
that are durable but not detectable.


\remove{
Friedman {\em et al.}~\cite{DBLP:conf/ppopp/FriedmanHMP18} presented
a detectably recoverable queue together with a set of abstract guidelines for building detectable and recoverable data structures. 
We followed these guidelines to come up with a linked-list implementation, called \DTOpt\, which we also compare with \ISB.
}



We have also undertaken the challenging task of adding persistency instructions in a manual, hand-tuned way
to both \ISB\ and \Capsules.
This resulted in two new implementations, called 
\ISBOpt\ and \CapsulesOpt, respectively. 
Moreover, we developed an additional technique, called \emph{direct tracking} (DT). 
Direct tracking uses an algorithmic idea utilized in~\cite{DBLP:conf/ppopp/FriedmanHMP18}
for implementing a recoverable queue.
It is applicable to implementations in which every update takes effect in \emph{a single} \CAS{} (e.g.,~\cite{RF2004,DBLP:conf/wdag/Harris01,MichaelS-PODC1996})
and requires an \emph{arbitration} mechanism that
helps determine the responses of updates that failed while
competing to apply the same change to the data structure
(e.g., deleting the same node).
Upon recovery, each of these processes competes to become the one to
which the successful execution of the primitive operation is attributed,
thereby determining its response value. 
\emph{Persistency instructions} (i.e., flushes and fences) were added 
based on the abstract guidelines provided in~\cite{DBLP:conf/ppopp/FriedmanHMP18}.
Doing so was not an easy task 
as~\cite{DBLP:conf/ppopp/FriedmanHMP18} does not provide a mechanical approach
for applying the guidelines in order to get a recoverable data structure.
The resulted implementation has been optimized in a hand-tuned way to further
reduce the persistency overhead. This resulted in \DTOpt. 

In addition to linked list algorithms, we also implemented an ISB-based queue and compared it against a capsules-based queue~\cite{BBFW2019} and a log queue~\cite{DBLP:conf/ppopp/FriedmanHMP18}. Evaluation results show that none of the queue algorithms scale and their throughput becomes very low for 16 or more threads. The ISB-based queue outperforms the two other algorithms for 8 or more threads. For lack of space, the results of our queue experiments are described in the supplementary material.

\noindent
{\bf Experimental setting and benchmarks.}
We used a 40-core machine with 4 Intel(R) Xeon(R) E5-4610 v3 1.7Ghz CPUs
with 10 cores each with hyper-threading support (for a total of 80 hardware threads) and 25MB L3 cache. The machine runs CentOS Linux 7.5.1804
with kernel version 3.10.0-862.14.4.el7.x86\_64 and has 256GB RAM.
Code is written in C++ and compiled using g++ (version 4.8.5) with O3 optimizations.
Each experiment lasts $5$ seconds and each data point is the average of $10$ experiments.
Keys are chosen uniformly at random from the ranges $[1, 500]$ or $[1, 1500]$.
Experiments with more key ranges can be found in the supplementary material, 
while they exhibit the same trends with our analysis below.
The list is initially populated by performing $250$ or $750$ \insertlst s,
respectively, resulting in an almost $40\%$-full list.
We present update-intensive (30\% finds) and
read-intensive (70\% finds) benchmarks.
Results for other 
operation type distributions were similar. 

As we do not have machines with NVRAM in our sites to run our experiments,
we follow a similar approach as in~\cite{BBFW2019,DBLP:conf/ppopp/FriedmanHMP18}
and simulate \pwb\ using \clflush\ and \psync\ using \mfence, expecting performance overhead
close to the
real overhead of a persistency
instruction~\cite{Bhandari:2016:MFR:3022671.2984019, Chakrabarti:2014:ALL:2714064.2660224}
in systems supporting NVRAM (such as 3DXPoint).
Since we assume TSO, there is no
need to simulate \pfence. Experimentation on machines with
NVRAM is left for future work.

\begin{figure*}
	\centering
	\begin{subfigure}{0.3\textwidth}
		\includegraphics[width=1\textwidth]{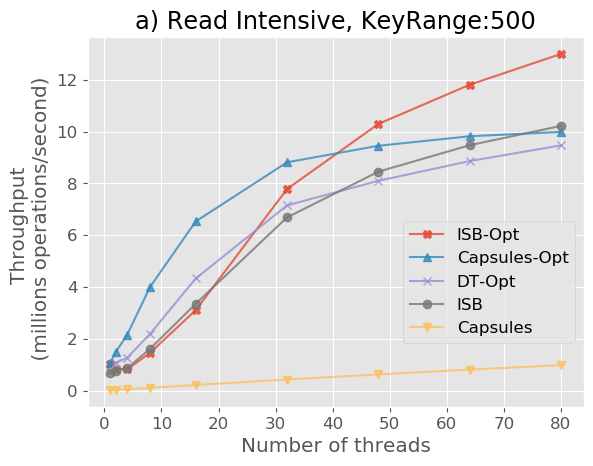}
		\phantomsubcaption
		\label{fig:evaluation:a}
	\end{subfigure}		
	\begin{subfigure}{0.3\textwidth}
		\includegraphics[width=1\textwidth]{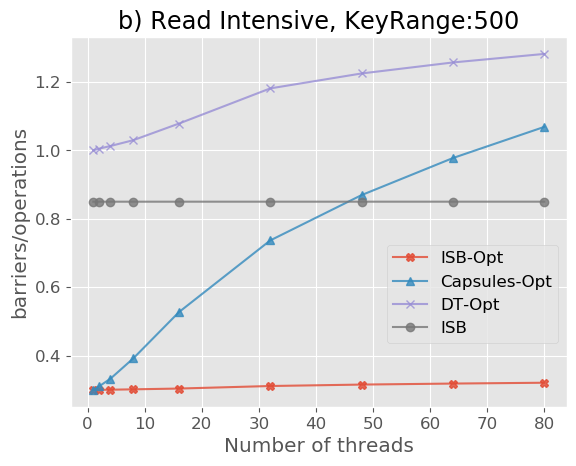}
		\phantomsubcaption
		\label{fig:evaluation:b}
	\end{subfigure}		
	\begin{subfigure}[c]{0.3\textwidth}	
		\includegraphics[width=1\textwidth]{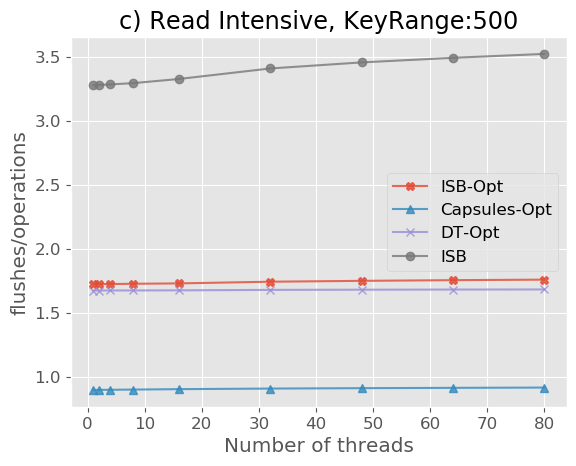}
		\phantomsubcaption
		\label{fig:evaluation:c}
	\end{subfigure}		
	\begin{subfigure}[c]{0.3\textwidth}	
		\vspace{-0.35cm}
		\includegraphics[width=1\textwidth]{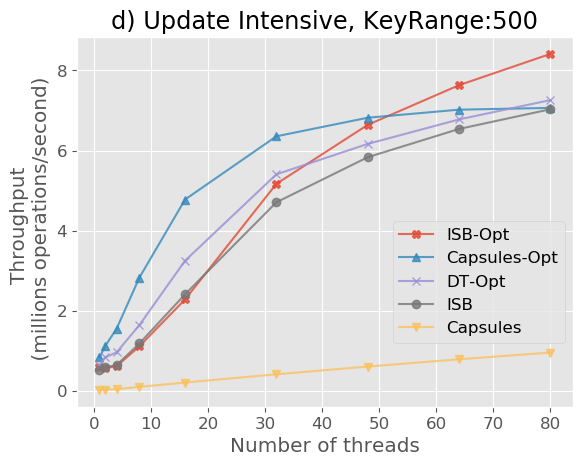}
		\phantomsubcaption
		\label{fig:evaluation:d}
	\end{subfigure}		
	\begin{subfigure}[c]{0.3\textwidth}	
		\vspace{-0.35cm}
		\includegraphics[width=1\textwidth]{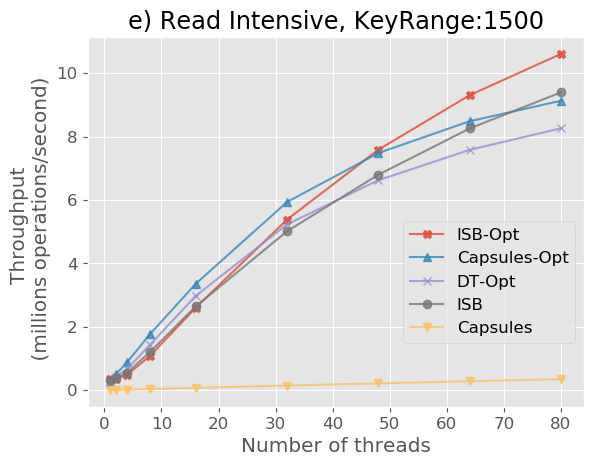}
		\phantomsubcaption
		\label{fig:evaluation:e}
	\end{subfigure}		
	\begin{subfigure}[c]{0.3\textwidth}	
		\vspace{-0.35cm}
		\includegraphics[width=1\textwidth]{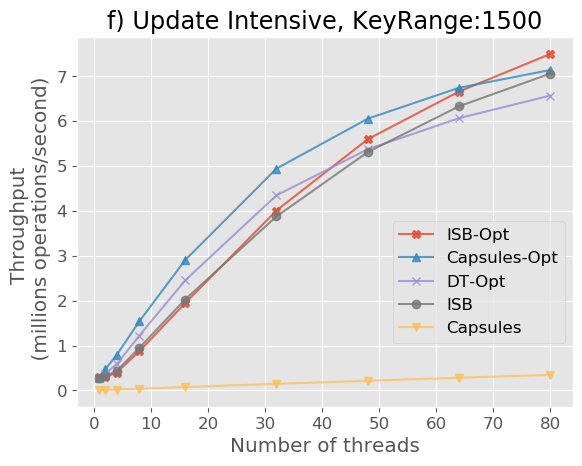}
		\phantomsubcaption
		\label{fig:evaluation:f}
	\end{subfigure}		
	\vspace{-0.5cm}
	\caption{Throughput and number of \barrier s\ and stand-alone flushes in evaluated implementations. }
	\label{fig:evaluation}
\end{figure*}

\noindent
{\bf Experimental Analysis.}
The results of our experiments are shown in Figure~\ref{fig:evaluation}.
Throughput evaluation results are shown by Figures~\ref{fig:evaluation}a, \ref{fig:evaluation}d, 
\ref{fig:evaluation}e and \ref{fig:evaluation}f.
The throughput of \CP{} 
is extremely low due  to the 
high number of persistency instructions incurred by the transformation in~\cite{DBLP:conf/wdag/IzraelevitzMS16}. 
On the contrary,
\ISB\ exhibits good performance despite its generality. 

We next compare the performance of \ISB\ with the linked list implementations
that have been optimized in a hand-tuned manner,
namely \CapsulesOpt\ and \DTOpt. 
As we see in Figure~\ref{fig:evaluation}, our general scheme has comparable 
(and sometimes better) performance to that of these algorithms. 
This shows that the overhead of \ISB\ is low. 

To achieve a fairer comparison,
we have developed a hand-tuned optimized version of
\ISB, called \ISBOpt, which we compare with \CapsulesOpt\ and \DTOpt.
It can be seen that \ISBOpt's relative performance 
improves as the number of threads increases.
Specifically, in the read-intensive case and for small key ranges
(that result in higher contention), the speedup of both \DTOpt\ and \CapsulesOpt\ 
becomes very small after 32 threads, whereas \ISBOpt\ continues to exhibit 
significant speedup up to the 80 supported threads.
%
\ISBOpt's scalability stems from the fact that it performs fewer barriers per operation (Figure~\ref{fig:evaluation}b). 

We note that if a node is deleted, its marked bit 
must be persisted. Otherwise, the following bad scenario
may happen: a thread executing \func{Find}, searching for
a key $k$ which has been logically deleted without persisting
its marked bit, may run to completion and return \False. 
Then, a crash may cause the logically deleted node to appear
in the linked list as unmarked. A subsequent \func{Find}
would then return \True, which is incorrect.
Because of the way helping is performed in ISB-tracking, 
no paths of marked nodes can be created in the data structure. 
Therefore, there are no chains of nodes that have changed and
need to be persisted. 
So, a thread has to
perform \barrier\ (i.e., a \clflush\ followed by an \mfence)
only on nodes in the \AffectSet\ of its current operation $Op$.
It thus performs a constant number of \barrier s in each attempt of $Op$,
regardless of the total number of threads (see Figure~\ref{fig:evaluation}b).
In contrast, in both \CapsulesOpt\ and \DTOpt, 
an operation may depend on a long chain of nodes that are yet to be persisted. 
So,
both \CapsulesOpt\ and \DTOpt\ perform a \barrier\
each  time they traverse a marked node. 
As the number of threads increases,
\barrier\ is performed on more marked nodes,
increasing the persistency cost.

We also see (Figure~\ref{fig:evaluation}c) that \ISBOpt\ performs on average more {\em stand-alone} \clflush\ instructions (not included in a \barrier)
than the other algorithms (but still only a constant number per operation).
This results from the need to persist $CP$, $RD$, and other
variables. 
This explains why \ISBOpt\ is outperformed by the other algorithms
when the number of threads is small, as well as why the crossing point
of the \ISBOpt\ curve with the other curves moves to the right
in update-intensive workloads. 
For large key ranges, the list is longer and
marked nodes are more scattered in the list, so a thread $q$
traverses more nodes between two marked nodes $nd_1$ and $nd_2$.
This often provides sufficient time for the thread that marked $nd_2$ to 
physically remove it from the list,
so that $q$ never reaches it, thus reducing the number of marked nodes traversed by threads.
Experimental results for the private-cache model,
where no algorithm incurs persistency cost,
support these observations (see supplementary material).


\remove{
The results \lefteris{of our experimental analysis for queues} (Figure~\ref{fig:ev-apnd-queues}) 
show that none of the queue algorithms is
scalable 
and their throughput becomes very low for 16 or more threads. 
The ISB-based queue outperforms the two other algorithms for 8 or more threads. 
Additionally, the ISB-based queue performs better than \CPG\ for 3 or more threads and better 
than log queue for 6 or more threads.

\begin{figure*}
	\centering
	\includegraphics[width=0.32\textwidth]{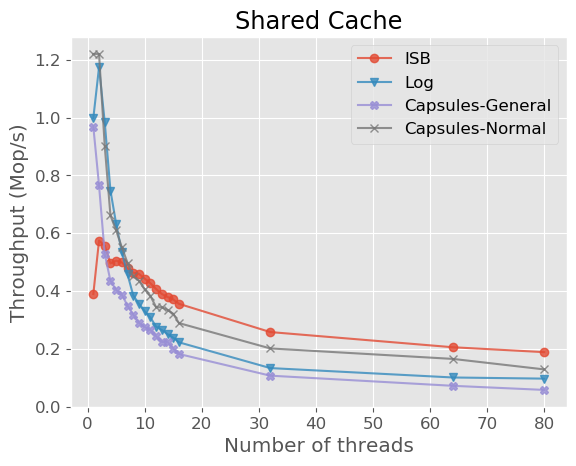}
	\includegraphics[width=0.32\textwidth]{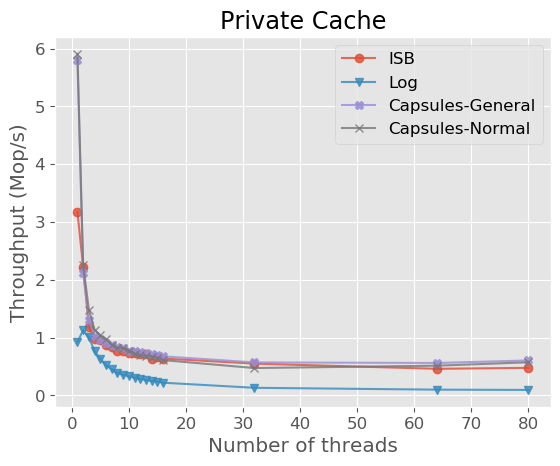}
	\includegraphics[width=0.32\textwidth]{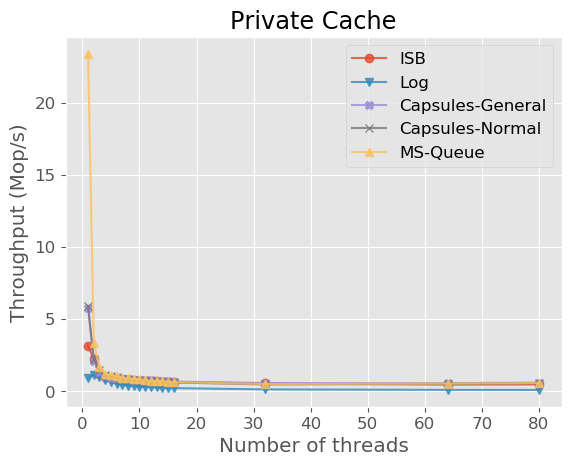}	
	\caption{Throughput of evaluated queue implementations in the shared (left) and the
	private (middle and right) cache models. The right figure is the same with the middle,
	while it also includes the throughput of \MSQUEUE.}
	\label{fig:ev-apnd-queues}
\end{figure*}

}

\section{Detectably Recoverable Versions of Additional Data Structures}
\label{section:elimination-stack}
\label{section:BST}
We briefly discuss additional data structures that can 
become detectably recoverable by applying the ISB approach. 
Due to lack of space, the details of the implementations discussed below will be provided in the paper's full version.

\noindent
{\bf Detectably Recoverable Binary Search Tree.}
The algorithm in~\cite{DBLP:conf/podc/EllenFRB10} (\NBBST) implements a {\em leaf-oriented} ({\em external}) binary search tree.
It uses \CAS\ to {\em flag} an internal node whenever a child pointer of it is to be changed,
and to {\em mark} it whenever it is to be deleted.
A process $p$, executing an update $Op$, allocates an Info structure
where it records the information
needed by other processes to help $Op$ complete.
Each internal node
contains an {\em update} field which stores a reference to an Info structure
and a 2-bits $status$ field
which 
indicates whether the node is {\em flagged} for insertion, {\em flagged} for deletion, {\em marked}, or clean.
Each successful flag or mark \CAS\  installs
a pointer to the Info structure of the relevant operation in the update field
of the node it is applied on.

We employ the ISB-tracking approach (Algorithm~\ref{alg:ROpt-ISB-generic}) 
to make \NBBST\ detectably recoverable.
Consider an operation $Op$ 
and let
$l$, $p$ and $gp$ be pointers to the leaf $Op$'s search arrives at, to its parent
and to its grandparent, respectively. 
If $Op$ is an \func{Insert}, it replaces the node pointed to by $l$ with 
a subtree of three nodes. 
Thus, $Op$'s \AffectSet\ contains
a pointer to $l$ and a pointer to $p$  (as its child pointer will change 
to point from $l$ to the root of the new subtree). 
$Op$'s \WriteSet\ contains $p$, and $Op$'s \NewSet\
contains the three new nodes of the subtree 
that replaces $l$.
If $Op$ is a \func{Delete}, $Op$ changes the appropriate child pointer of $gp$ 
to point to the sibling of $l$. For applying ISB-tracking, we need to create
a copy of this sibling, to avoid the ABA problem. Therefore,
$Op$'s \AffectSet\ contains $l$, $p$, $gp$,
and a pointer to $l$'s sibling;
its \WriteSet\ contains $gp$ and its \NewSet\ 
the new node that replaces $l$'s sibling. 
The \AffectSet\ of a \func{Find} contains only $l$. 
\func{Find}s can be further optimized 
to have their \AffectSet\ be equal to the empty set.
Processes can use the $update$ field that already
exists in the tree nodes and the Info structures used in \NBBST, 
to implement ISB-tracking without any significant memory overhead. 
Also, the tagging mechanism is provided for free through the flagging and marking mechanism of \NBBST.

\noindent
{\bf Detectably Recoverable Exchanger.}
An \emph{Exchanger}~\cite{DBLP:books/daglib/0020056,scherer2005scalable}
allows two processes to pair-up the operations they are executing
and exchange values.
The first process, $p$, to arrive to an Exchanger, finds it {\em free}
and {\em captures} it by
atomically writing to it its value. 
Then, $p$ busy-waits until another process $q$ \emph{collides} with it:
if $q$ arrives while $p$ is waiting,
it reads $p$'s value in the Exchanger, and tries 
to atomically write its value to it 
and inform $p$ of a successful collision.

We employ the tracking approach to achieve recoverability:
processes exchange Info structures (ExInfo) instead of values.
In addition to \emph{state} 
and \emph{value} fields, \exInfo\ contains a \emph{result} field,
and a \emph{partner} field pointing to the ExInfo of the operation
with which $p$ is trying to collide. 

\section{Related Work and Discussion}
\label{section:related}

We present the ISB-tracking approach for detectable recovery of
concurrent data structures and
apply it to many well-known concurrent data structures,
e.g., queues, linked lists, trees, as well as to exchangers.
There are several recoverable concurrent implementations of specific
data structures such as
mutual exclusion locks~\cite{DBLP:conf/podc/GolabH17,GolabR16},
queues~\cite{DBLP:conf/ppopp/FriedmanHMP18,DBLP:conf/wdag/JayantiJ17}
and B-trees~\cite{chi2014making,VenkataramanTRC-FAST2011,moraru2012persistent},
with optimizations exploiting \emph{specific} aspects of the objects.
In contrast, our approach is general and derives recoverable implementations
from their non-recoverable counterparts,
preserving their efficiency.

Our implementations are \emph{strictly
recoverable}~\cite{AttiyaBH-PODC2018}: 
the response of a recoverable operation $Op$ is made persistent
before $Op$ completes, so that a higher-level operation that
invokes $Op$ is able to access $Op$'s response value,
even if a crash occurs after $Op$ completes.
They also satisfy \emph{nesting-safe recoverable linearizability}
(NRL)~\cite{AttiyaBH-PODC2018}: %
a failed operation is linearized within an interval
that includes its failures and recovery attempts.
This implies \emph{durable linearizability}
(DL)~\cite{DBLP:conf/wdag/IzraelevitzMS16}---the
state of the object after a crash reflects a consistent operation
sub-history including all operations completed by the time of the crash,
and some operations in progress when the crash occurred may be lost.
Info structures were used in DL implementations of several data
structures~\cite{CoburnCAGGJW-Asplos2011,
DBLP:conf/podc/PavlovicKMH18, DBLP:conf/icde/WangLL18},
and other transformations that avoid
logging~\cite{DBLP:conf/usenix/DavidDGZ18, DBLP:conf/wdag/IzraelevitzMS16},
but none of them ensures detectability.

The recoverable \emph{log queue}~\cite{DBLP:conf/ppopp/FriedmanHMP18}
augments queue nodes with tracking information, which is used
after a system-wide crash to \emph{synchronously} try and complete
all pending operations 
from the previous phase before starting a new phase.
Two other recoverable queues in~\cite{DBLP:conf/ppopp/FriedmanHMP18}
are not detectable. 

An NRL implementation can be obtained from any algorithm
using only read, write and \CAS\ primitives
by replacing each primitive with its (NRL) recoverable version
(see~\cite{AttiyaBH-PODC2018}).
Implementations using only read and \CAS\ can be made
recoverable and detectable
using capsules~\cite{BBFW2019} 
(see Section~\ref{section:evaluation}).
\here{HA: Removed the following, because we have evaluation.}
\here{HA: Removed the following, because we have evaluation.}

A recoverable lock-free universal implementation~\cite{DBLP:conf/spaa/CohenGZ18}
requires only one round trip to NVRAM per operation, which is optimal.
It is essentially log-based,
keeping the entire history of the object in a designated shared queue.
It also keeps a per-process persistent log, such that, collectively,
these logs keep the entire history, but different logs may have a big overlap.
To determine the response of an operation,
the entire history is read until its linearization point,
where the operation's response can be determined,
making the construction detectable.
This construction 
makes the strong assumption that a single recovery function is
executed upon recovery, consistently reconstructing the data structure,
whereas we allow failed processes to be recovered
by the system in an asynchronous manner.
Romulus \cite{DBLP:conf/spaa/CorreiaFR18} is a
persistent transactional memory framework that provides durability and detectability.
However, it is blocking, satisfying only starvation-freedom for update transactions.
Other logging-based approaches
are~\cite{chatzistergiou2015rewind,DBLP:journals/pacmpl/CohenFL17}.

Our observations that helping can be leveraged for recovery resemble
the usage of re-entrant and idempotent helping for lock-free memory
reclamation in lock-free data structures~\cite{DBLP:conf/podc/Brown15}.

Our recoverable implementations---as well as the original, non-recoverable
implementations---rely on garbage collectors that correctly
recycle memory once it becomes unreachable.
This naturally motivates the question of implementing lock-free recoverable
memory managers~\cite{Bhandari:2016:MFR:3022671.2984019,RamalheteCFC19}, which we plan to investigate in future work.
Another research avenue we plan to pursue is further experimental evaluation
of our recoverable implementations, in particular,
the interaction of the NVRAM with system caches.


\newpage



\bibliographystyle{abbrv}
\bibliography{references}

\appendix
\newpage
\clearpage
\begin{center}
\LARGE{\bf{ Supplementary Material for Submission \#215} \\
Tracking in Order to Recover: Detectable Recovery of Lock-Free Data Structures}
\end{center}
\section{Detectably Recoverable Linked List - Additional Code}
\label{section:apendix-LL}

In this section, we provide additional material for the
detectably recoverable linked list implementation presented 
in Section~\ref{section:linked-list}, including data types,
shared variables, and initialization values in Figure~\ref{alg:ROpt-ISB-ll-definition}
and the pseudocode for \func{Delete} in Algorithm~\ref{alg:ROpt-ISB-ll-delete}.

\begin{figure}[b]
	
		\scriptsize
		
		\SetKwBlock{Begin}{}{}		
		
		\begin{flushleft}
			
		type Node \{ \hspace*{13.25mm}  \\
		\hspace*{6mm} $\Key \cup \{\infty, -\infty\}$ $key$ \\
		\hspace*{6mm} Node *$next$ \\
		\hspace*{6mm} Info *$info$ \\
		\}

		type \Info\ \{ \\
		\hspace*{6mm} \{\delete, \insertlst, \find \} opType \\
		\hspace*{6mm} Set \AffectSet; \\
		\hspace*{6mm} Set \WriteSet; \\
		\hspace*{6mm} Set \NewSet; \\		
		\hspace*{6mm} \recovercode{boolean $result$} \\
		\}

		\com Initialization: \\
		\hspace*{6mm} Shared Node: *$head$ with key $-\infty$, *$tail$ with key $\infty$. \\
		\hspace*{6mm} $head \rightarrow next$ points to $tail$,  $tail \rightarrow next$ points to \NULL. \\
		\hspace*{6mm} Both $info$ fields are \NULL. \\
		
		\end{flushleft}
	\caption{Recoverable Linked List types and initialization.}
	\label{alg:ROpt-ISB-ll-definition}
\end{figure}

\begin{algorithm}[tb]	
	\nonl
	
		
		\removelatexerror
		\scriptsize
		
		\SetKwBlock{Begin}{}{}			
		
		\begin{procedure}[H]
			\caption{() \small $\langle$Node$^*$, Node$^*$, Info$^*$, Info$^*$$\rangle$ \search\ (T $key$)}
			
			Node *\textit{pred}, *\textit{curr} \;
			Info *\textit{predInfo}, *\textit{currInfo} \;
			
			\textit{curr} := \textit{head} \;
			\textit{currInfo} := \textit{head }$\rightarrow$\textit{info} \;
			\tcc{Search for first node with key $\geq$ \textit{key}}
			\While{\textit{curr}$\rightarrow$\textit{key} $<$ \textit{key}}{
				$pred := curr$ \tcp*{remember predecessor}
				$predInfo := currInfo$ \tcp*{remember $info$ field of predecessor}
				$curr := curr \rightarrow next$ \tcp*{move forward in the list}
				$currInfo := curr\rightarrow info$ \tcp*{copy $info$ field}
			}
			\KwRet $\langle pred,curr, predInfo,currInfo \rangle$
		\end{procedure}
		
		\begin{procedure}[H]
			\color {blue} {
				\caption{() \small boolean \func{Op-Recover} (T $key$)}
				
				\Info\ *\textit{opInfo} $:= RD_q$ \label{ISB-ll-Template2-OpRecover-read-RDp} \;
				
				\uIf {$CP_q = 0$ \OR \textit{opInfo} $= \init$} {
					Re-invoke \func{Op} \label{ISB-ll-Template2-OpRecover-first-reinvoke}
				}


				\func{Help}(\textit{opinfo}, \True)

				\color {blue} {\uIf (\tcp*[f]{operation completed}) {\textit{opInfo} $\rightarrow$ \textit{result} $\neq \init$}
				{\KwRet \textit{opInfo} $\rightarrow$ \textit{result} \label{ISB-ll-Template2-OpRecover-return-true}}
				\lElse {
					Re-invoke \func{Op} \tcp*[f]{operation attempt failed} \label{ISB-ll-Template2-OpRecover-reinvoke}
				}}
			}
		\end{procedure}


	\begin{procedure}[H]
		\caption{() boolean \delete\ (T $key$)}
		
		Info *\textit{opInfo} $:=$ \textbf{new} \Info\ () \;
		
		\recovercode{$RD_q := \NULL$} \label{ISB-ll-Template2-delete-set-RD-init} \;
		\flushcode{\pbarrier\ ($RD_q$)} \;
		\recovercode{$\text{CP}_q := 1$} \tcp*{check-point; $RD_q$ is initialized} \label{ISB-ll-Template2-delete-set-checkpoint}
		\flushcode{\pwb\ ($\text{CP}_q$)}; \flushcode{\psync} \;
		
		\While{\True}{
			\Begin (\textbf{Gather Phase} \tcp*[f]{search for node to delete})  {
				$\langle$\textit{pred}, \textit{curr}, \textit{predInfo}, \textit{currInfo}$\rangle :=$ \search(\textit{key}) \label{ISB-ll-Template2-delete-search} \;
				\lIf {\textit{curr}$\rightarrow$\textit{key} $\neq$ \textit{key}} {
					\AffectSet\ $:= \{\langle$\textit{curr},\textit{currInfo}$\rangle\}$
				}
				\lElse {
					\AffectSet\ $:= \{\langle$\textit{pred},\textit{predInfo}$\rangle,\langle$\textit{curr},\textit{currInfo}$\rangle\}$
				}
			}
			\Begin (\textbf{Helping Phase} \tcp*[f]{help other operations if necessary}) {
				\lIf {\textit{isTagged}(\textit{predInfo})} {\func{Help}(\textit{predInfo}) \label{ISB-ll-Template2-delete-help-pred}}
				\lElseIf {\textit{isTagged}(\textit{currInfo}} {\func{Help}(\textit{currInfo}) \label{ISB-ll-Template2-delete-help-curr}}
				\lIf {\textit{isTagged}(\textit{predInfo}) \OR \textit{isTagged}(\textit{currInfo})} { \Continue }
				
			}
			\WriteSet := $\{\langle$\textit{pred}$\rightarrow$\textit{next}, \textit{curr}, \textit{curr} $\rightarrow$ \textit{next} $\rangle\}$ \;
			*\textit{opInfo} $:= \langle$\delete, \AffectSet, \WriteSet, $\emptyset$, $\init \rangle$ \label{ISB-ll-Template2-delete-create-info} \;

			\yyy{\uIf (\tcp*[f]{\textit{key} not in the list}) {\textit{curr} $\rightarrow$ \textit{key} $\neq$ \textit{key} \label{ISB-ll-Template2-delete-if-in-list}} {
				\recovercode{\textit{opInfo} $\rightarrow$ \textit{result} $:=$ \False} \;
			}}
			\flushcode{\pbarrier\ (\textit{opInfo})} \;
			\recovercode{$RD_q :=$ \textit{opInfo}} \tcp*{info for current attempt} \label{ISB-ll-Template2-delete-update-RD}
			\flushcode{\pwb\ ($RD_q$)}; \flushcode{\psync} \;
			\yyy{\uIf (\tcp*[f]{\textit{key} not in the list}) {\textit{curr} $\rightarrow$ \textit{key} $\neq$ \textit{key} \label{ISB-ll-Template2-delete-if-in-list}} {
				\KwRet \False \label{ISB-ll-Template2-delete-return-false}
			}}
			\func{Help}(\textit{opInfo}, \True) \;
			\lIf {\textit{opInfo} $\rightarrow$ \textit{result} $\neq \init$} {
				\KwRet \textit{opInfo} $\rightarrow$ \textit{result}
			}
		}
	\end{procedure}


	\caption{Recoverable Linked List \func{Delete}, \func{Op-Recover} and auxiliary functions (code for process $q$)}
	\label{alg:ROpt-ISB-ll-delete} \label{alg:ISB-ll-Template2-search}
\end{algorithm}

\section{Additional Experimental Results}
\label{section:appendix-evaluation}

\subsection{Linked Lists}
\label{section:appendix-evaluation:LL}
\begin{figure*}
	\centering
		\includegraphics[width=0.4\textwidth]{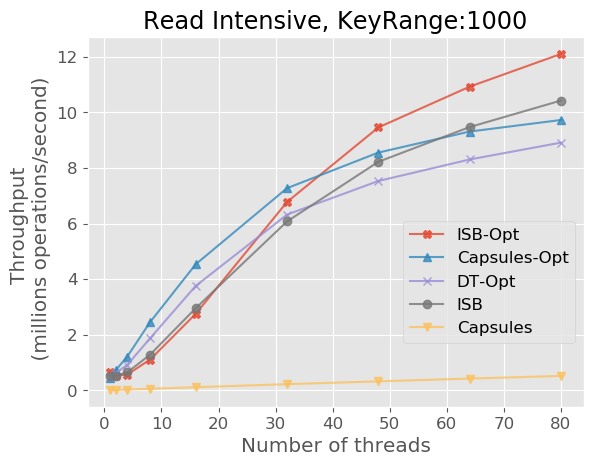}
		\includegraphics[width=0.4\textwidth]{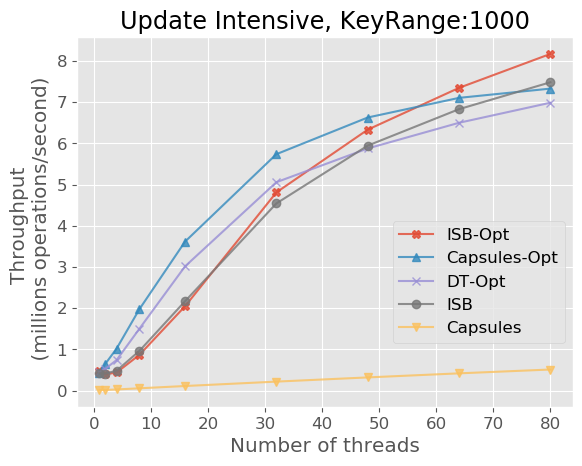}
		\includegraphics[width=0.37\textwidth]{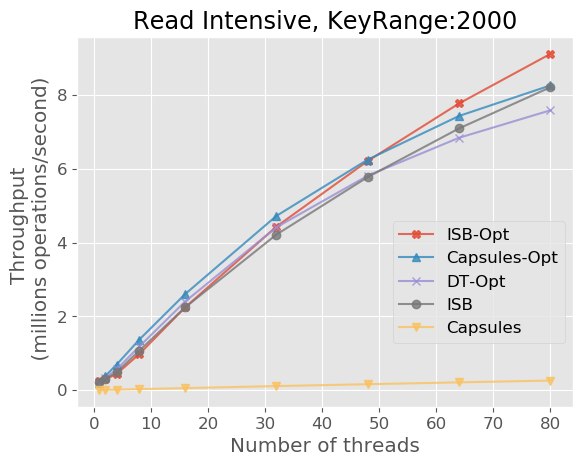}
		\includegraphics[width=0.37\textwidth]{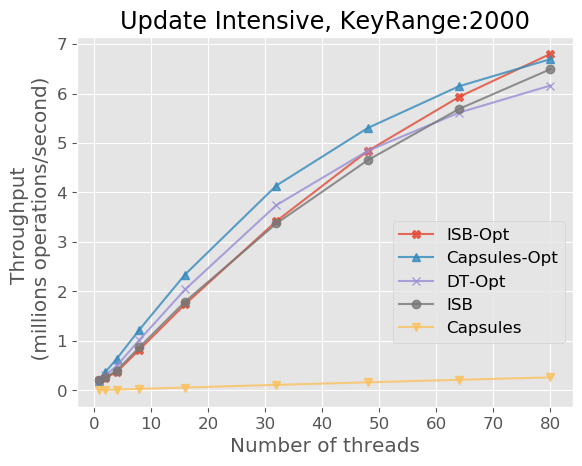}
	\caption{Throughput for key ranges [1,1000] and [1,2000] (the read-intensive benchmarks are
	shown on the left and the update-intensive benchmarks are shown on the right). }
	\label{fig:ev-apnd-1000-2000}
\end{figure*}

Figure~\ref{fig:ev-apnd-1000-2000} illustrates the throughput of all
linked list algorithms for the following key ranges: [1,1000], [1,2000]. 
Results for the read-intensive workload are presented on the left
and those for the update-intensive workload are presented on the right.
Throughput results exhibit the same trends we described and analyzed in Section~\ref{section:evaluation} for other key ranges.

\begin{figure*}
	\centering
		\includegraphics[width=0.4\textwidth]{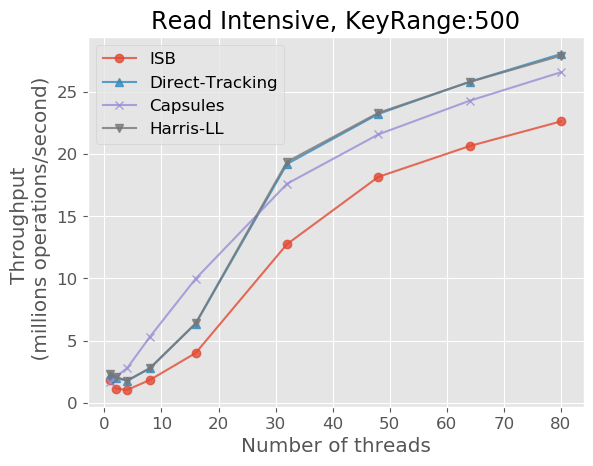}
		\includegraphics[width=0.4\textwidth]{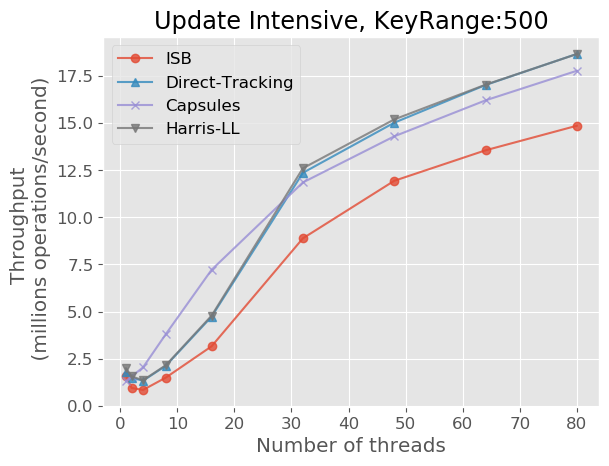}
	\caption{Algorithms' throughput in the private cache model. }
	\label{fig:ev-apnd-pc}
\end{figure*}

Figure~\ref{fig:ev-apnd-pc} presents the performance of the evaluated algorithms in the private cache model,
where the persistency cost is zero (since no flushes or fences are required). This experiment emphasizes the 
overheads that are incurred by the algorithms due to additional metadata maintenance and \CAS\ operations 
performed for guaranteeing detectable recoverability.
In addition to evaluating the ISB-based, capsules and direct tracking linked lists, we 
also evaluate the original (\emph{non-recoverable}) linked list implementation 
of Tim Harris~\cite{DBLP:conf/wdag/Harris01}, which we name \HarrisLL.
All algorithms exhibit speedup, when there is no persistency cost.
As expected, the performance of the \HarrisLL\ algorithm is almost identical to 
that of direct tracking since the latter is based on the former and mainly adds 
to it persistency instructions required for detectability, which are not performed in 
the private cache model evaluation. These two algorithms outperform the capsules 
algorithm for 32 threads or more, but are outperformed by it for 
smaller numbers of threads. Note, however, that while the \HarrisLL\ and the direct 
tracking algorithms were handcrafted, the capsules and ISB-based linked list are the 
results of general transformations.

The performance of \ISB\ lags behind that of capsules by up to 5\% due to 
the overhead for maintaining the info structures. As we have shown in 
Section~\ref{section:evaluation}, this gap is decreased and even 
reversed for large numbers of threads, when persistence instructions 
are performed, as explained by our analysis in Section~\ref{section:evaluation}.

\begin{figure*}
	\centering
		\includegraphics[width=0.32\textwidth]{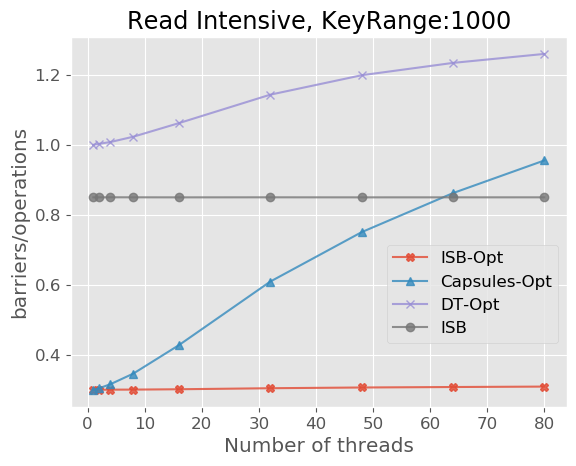}
		\includegraphics[width=0.32\textwidth]{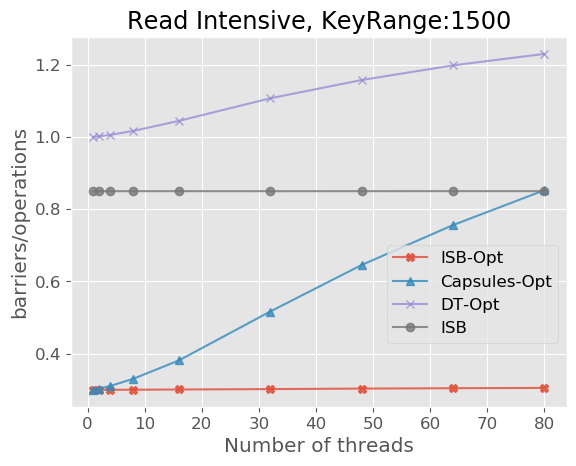}
		\includegraphics[width=0.32\textwidth]{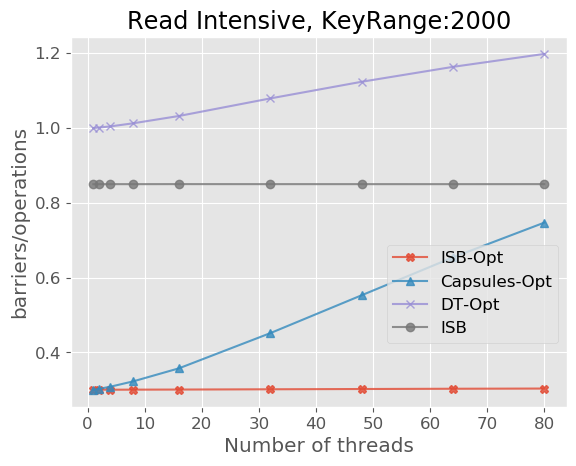}
		\includegraphics[width=0.32\textwidth]{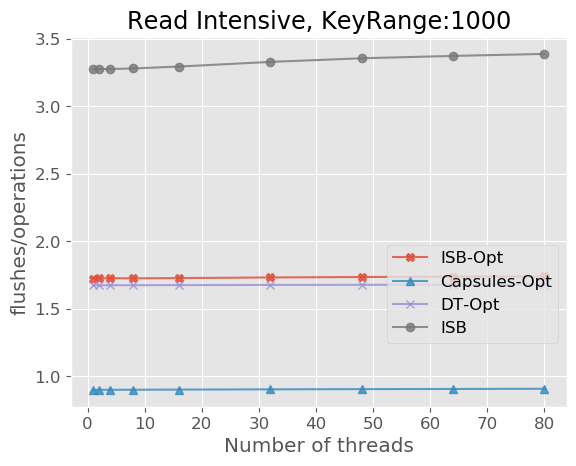}
		\includegraphics[width=0.32\textwidth]{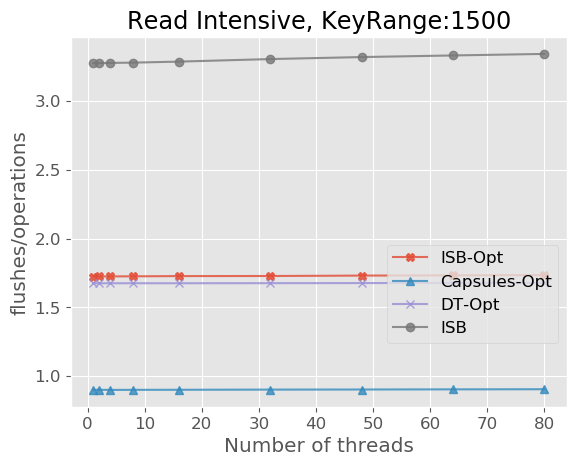}
		\includegraphics[width=0.32\textwidth]{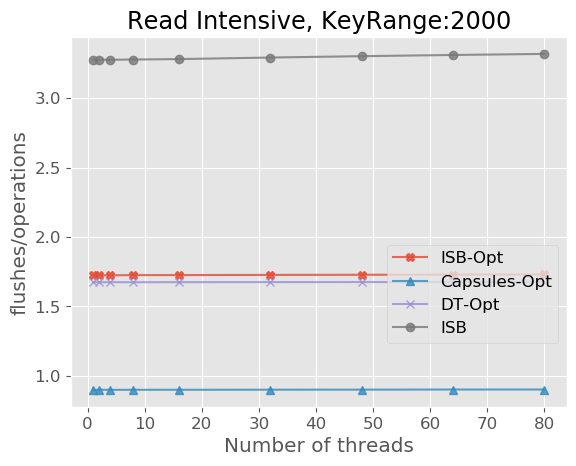}
	\caption{Number of \pbarrier s and stand-alone flushes performed by the different algorithm in read-intensive benchmarks for different key ranges (the number of \pbarrier s are shown on top, and the number of flushes are shown at the bottom). }
	\label{fig:ev-apnd-barrier-1}
\end{figure*}

\begin{figure*}
	\centering
		\includegraphics[width=0.32\textwidth]{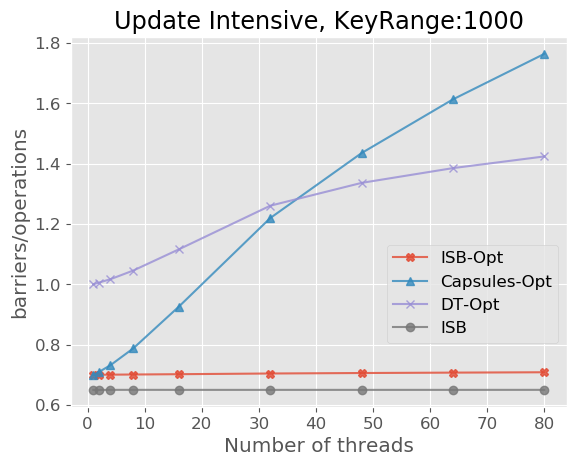}
		\includegraphics[width=0.32\textwidth]{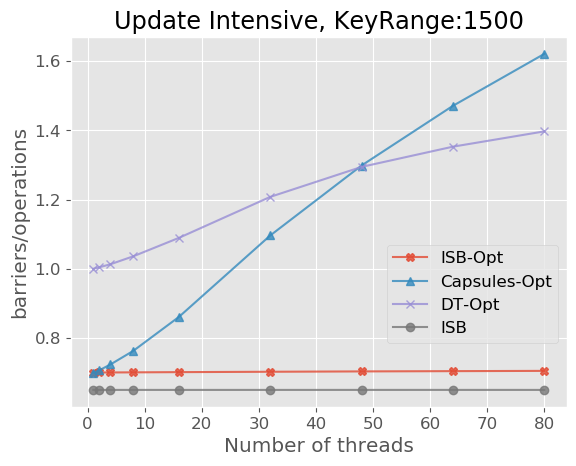}
		\includegraphics[width=0.32\textwidth]{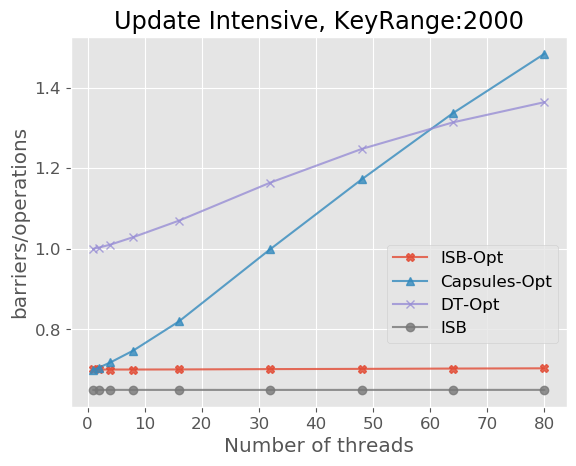}
		\includegraphics[width=0.32\textwidth]{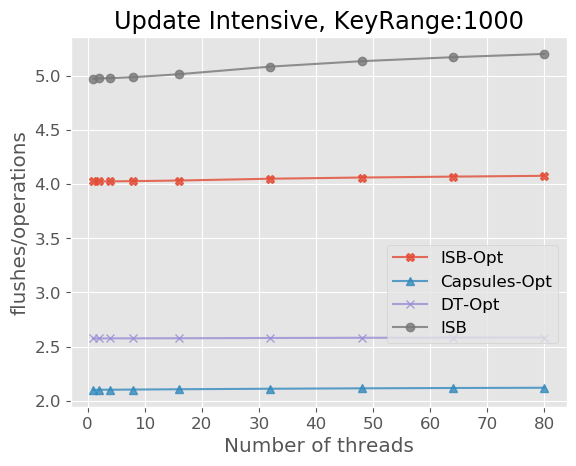}
		\includegraphics[width=0.32\textwidth]{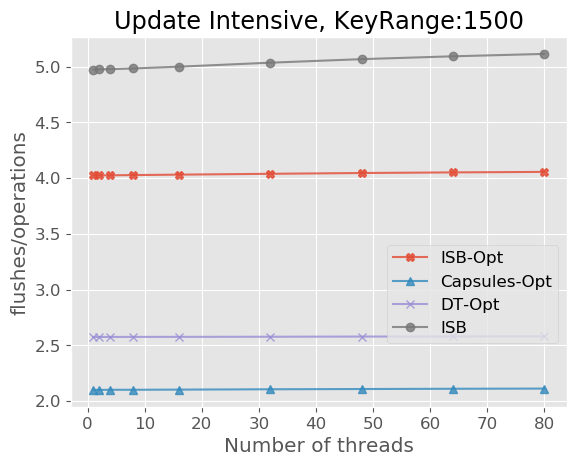}
		\includegraphics[width=0.32\textwidth]{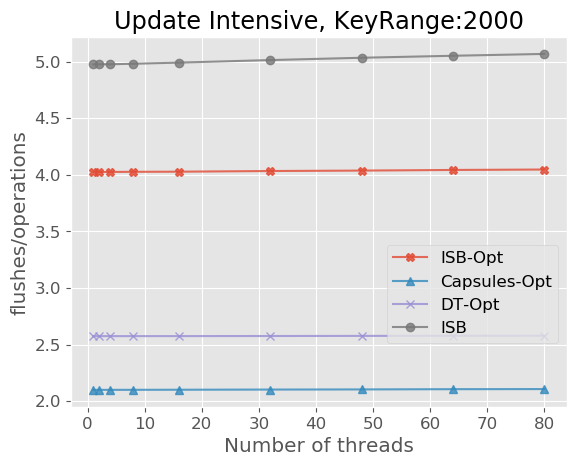}		
	\caption{Number of \pbarrier s and stand-alone flushes performed by the different algorithm in update-intensive benchmarks for different key ranges (the number of \pbarrier s are shown on top, and the number of flushes are shown at the bottom).}
	\label{fig:ev-apnd-barrier-2}
\end{figure*}

Figure~\ref{fig:ev-apnd-barrier-1} and~\ref{fig:ev-apnd-barrier-2} show the numbers of \pbarrier s and stand-alone \clflush\ instructions
for the read-intensive and the update-intensive benchmarks, respectively,
and key ranges: [1,1000], [1,1500], [1,2000]. They show the same trends exhibited by the experiments described in Section~\ref{section:evaluation}.

\subsection{Queues}
\label{section:appendix-evaluation:Queues}
We also implemented an ISB-based queue, 
after adding flushes and fences as described in Section~\ref{section:overview},
and compared it with two other recoverable and detectable queues:
the log queue~\cite{DBLP:conf/ppopp/FriedmanHMP18}
and the capsules-based queue~\cite{BBFW2019},
obtained by applying the capsules transformation
to \MSQUEUE~\cite{MichaelS-PODC1996}.
Specifically, we compared our ISB-based queue with 
two 
variants resulting from the capsules transformation:
\CPG{} applies the capsules transformation to any data structure
implementation, while \CPN{} applies a version of the capsules transformation
that is optimized for normalized implementations~\cite{TimnatP-PPoPP2014},
since the corresponding queue implementation is normalized.
Persistency instructions were added to these implementations and were hand-tuned.
%
The implementations of the log queue and the two capsule-based queues
were provided by their authors.

As in~\cite{BBFW2019,DBLP:conf/ppopp/FriedmanHMP18,MichaelS-PODC1996},
each thread applies pairs of enqueue and dequeue operations.
Each experiment runs over $5$ seconds and each
data point is the average of $10$ experiments.
The queue is initially populated with one million nodes.
As in~\cite{BBFW2019,DBLP:conf/ppopp/FriedmanHMP18},
we simulate \pwb\ using \clflush\ and \psync\ using \mfence,
expecting performance overhead close to the
real performance overhead of a persistency
instruction~\cite{Bhandari:2016:MFR:3022671.2984019, Chakrabarti:2014:ALL:2714064.2660224}
in systems supporting NVM (such as 3DXPoint).
Since we assume TSO, there is no need to simulate \pfence.


The results (Figure~\ref{fig:ev-apnd-queues}) show that no algorithm is
scalable 
and the throughput of the ISB-based queue exceeds that of the other algorithms for 8 threads or more. 
Additionally, the ISB-based queue performs better than \CPG\ for 3 or more threads and better 
than log queue for 6 or more threads.

\begin{figure*}
	\centering
	\includegraphics[width=0.32\textwidth]{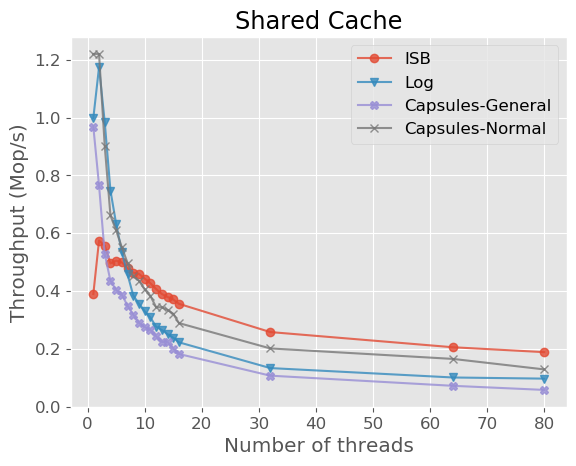}
	\includegraphics[width=0.32\textwidth]{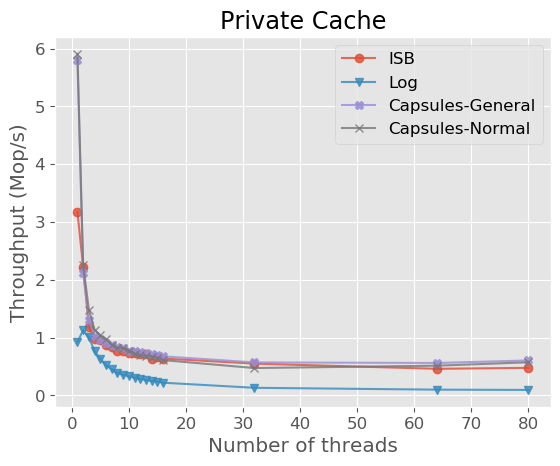}
	\includegraphics[width=0.32\textwidth]{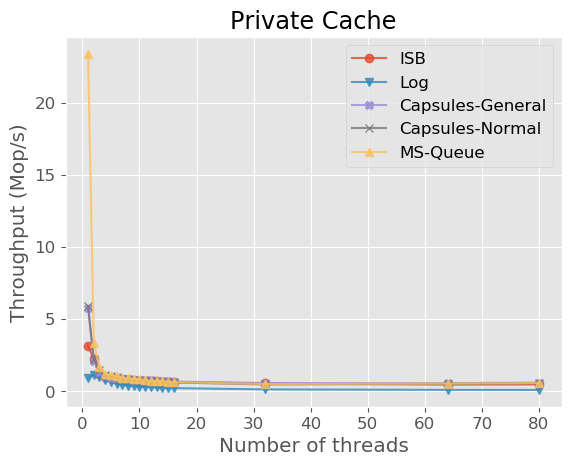}	
	\caption{Throughput of evaluated queue implementations in the shared (left) and the
	private (middle and right) cache models. The right figure is the same with the middle,
	while it also includes the throughput of \MSQUEUE.}
	\label{fig:ev-apnd-queues}
\end{figure*}




\end{document}